\documentclass[%
 reprint,
 superscriptaddress,
showpacs,
 amsmath,amssymb,
 aps,prc,
floatfix,
]{revtex4-1}

\usepackage{siunitx}
\usepackage{graphicx}
\usepackage{dcolumn}
\usepackage{bm}
\usepackage{wrapfig}
\usepackage{rotating}
\usepackage{epstopdf}
\usepackage{gensymb}

\newcommand*\addbar[1]{\overline{#1}}

\usepackage{silence}
\usepackage{xcolor} 

\begin{document}

\title{Decay spectroscopy of $^{50}$Sc and $^{50m}$Sc to $^{50}$Ti}

\author{M.~Bowry}
 \email{Michael.Bowry@uws.ac.uk}
\altaffiliation{Present address: School of Engineering, Computing and Physical Sciences, University of the West of Scotland, High Street, Paisley PA1 2BE, United Kingdom}
\affiliation{TRIUMF, 4004 Wesbrook Mall, Vancouver, BC, V6T 2A3, Canada}

\author{C.E.~Jones} 
\affiliation{TRIUMF, 4004 Wesbrook Mall, Vancouver, BC, V6T 2A3, Canada}
\affiliation{Department of Physics, University of Surrey, Guildford, Surrey, GU2 7XH, United Kingdom}

\author{A.B.~Garnsworthy}
 \affiliation{TRIUMF, 4004 Wesbrook Mall, Vancouver, BC, V6T 2A3, Canada}
 
\author{G.C.~Ball} 
\affiliation{TRIUMF, 4004 Wesbrook Mall, Vancouver, BC, V6T 2A3, Canada}

\author{S.~Cruz} 
\affiliation{Department of Physics and Astronomy, University of British Columbia, Vancouver, BC V6T 1Z4, Canada}
\affiliation{TRIUMF, 4004 Wesbrook Mall, Vancouver, BC, V6T 2A3, Canada}

\author{S.~Georges} 
\affiliation{TRIUMF, 4004 Wesbrook Mall, Vancouver, BC, V6T 2A3, Canada}

\author{G.~Hackman}
\affiliation{TRIUMF, 4004 Wesbrook Mall, Vancouver, BC, V6T 2A3, Canada}

\author{J.D.~Holt} 
\affiliation{TRIUMF, 4004 Wesbrook Mall, Vancouver, BC, V6T 2A3, Canada}

\author{J.~Measures} 
\affiliation{TRIUMF, 4004 Wesbrook Mall, Vancouver, BC, V6T 2A3, Canada}
\affiliation{Department of Physics, University of Surrey, Guildford, Surrey, GU2 7XH, United Kingdom}

\author{B.~Olaizola} 
\affiliation{TRIUMF, 4004 Wesbrook Mall, Vancouver, BC, V6T 2A3, Canada}

\author{H.P.~Patel} 
\affiliation{TRIUMF, 4004 Wesbrook Mall, Vancouver, BC, V6T 2A3, Canada}

\author{C.J.~Pearson} 
\affiliation{TRIUMF, 4004 Wesbrook Mall, Vancouver, BC, V6T 2A3, Canada}

\author{C.E.~Svensson}
\affiliation{Department of Physics, University of Guelph, Guelph, ON, N1G 2W1, Canada}

\date{\today}

\begin{abstract}
The $\beta$ decay of the isomeric and ground state of $^{50}$Sc to the semi-magic nucleus $^{50}_{22}$Ti$_{28}$ has been studied using a $^{50}$Ca beam delivered to the GRIFFIN $\gamma$-ray spectrometer at the TRIUMF-ISAC facility. $\beta$-decay branching ratios are reported to 16 excited states with a total of 38 $\gamma$-ray transitions linking them. These new data significantly expand the information available over previous studies. 
Relative intensities are measured to less than 0.001{\it $\%$} that of the strongest transition with the majority of $\gamma$-ray transitions observed here in $\beta$ decay for the first time. 
The data are compared to shell-model calculations utilizing both phenomenologically-derived interactions employed in the {\it pf} shell as well as a state-of-the-art, {\it ab initio} based interaction built in the valence-space in-medium similarity renormalization group framework.
\end{abstract}

\maketitle


\section{\label{sec:intro}Introduction}

Nuclei in the vicinity of magic neutron ({\it N}) and proton ({\it Z}) numbers, often display simple patterns of low-energy excitations which can be well described in a spherical shell model approach.
The structure of these lowest-lying excited states may be deduced by considering the behavior of a single nucleon or pair of nucleons occupying just a few single-particle orbits near the Fermi surface in a spherically-symmetric potential. However, this simple picture does not describe the nature of all excitations observed at low energies. Most notably, the limitations in the number of basis states included in this approach means that it usually does not capture deformed configurations or collective behaviors which can coexist along side the spherical single-particle structures and typically involve breaking the core. Particle-hole excitations across major shell gaps are energetically costly, requiring several MeV of energy. However, the additional correlation energy, coming primarily from the neutron-proton quadrupole-quadrupole interaction \cite{Heyde1982}, which becomes possible with this release of particles from the core makes such cross-shell excitations energetically competitive with the lowest-lying states \cite{Heyde1982,Heyde2011}.

Recently, the rapid development of new theoretical methods and the availability of increased computational power have extended the reach of {\it ab initio} methods to medium-mass nuclei \cite{Epel09RMP,Mach11PR,Bogn07SRG,Bogn10PPNP}. The need to include three-nucleon forces in the interactions for an accurate description of excitations has become evident \cite{Hebe15ARNPS}. In order to support the further development of these methods, detailed spectroscopic information of excited nuclear states and transitions is necessary.


The even-even {\it N}=28 isotones above $^{48}$Ca provide a good example, where protons fill the $0f_{7/2}$ orbital.
The seniority=2 ({\it V=2}), {\it J=$2^+$,$4^+$,$6^+$} states in $^{50}_{22}$Ti, $^{52}_{24}$Cr and $^{54}_{26}$Fe with {\it E$_x$} $\approx$1.5-3\,MeV show very little change in excitation energy as additional pairs of protons are added to the $0f_{7/2}$ orbital. In contrast the {\it J=$0_2^+$,$2_2^+$} excited states originating from neutron two-particle, two-hole ({\it $\nu$2p-2h}) excitations across the {\it N}=28 shell gap to the $1p_{3/2}$ orbital show an abrupt change beyond $^{50}$Ti. There is a rapid lowering in excitation energy as more proton pairs are added to increase the attractive correlation strength, and these states begin to intrude upon the $\pi 0f_{7/2}$ ground state structure at $\leq$ 3\,MeV in $^{52}$Cr and $^{54}$Fe \cite{rowanwood}. 
In addition, below $^{48}$Ca, the {\it N}=28 shell gap has been shown to vanish upon the significant removal of protons \cite{Sorlin12}.

The electric quadrupole transition strength {\it B(E2; $0_1^+$\,$\rightarrow$\,$2_1^+$)} is frequently used to probe the evolution of collectivity near closed shells, appearing enhanced at mid-shell and at a minimum at the shell closure, for example the {\it B(E2)} for the $2^+_1 \rightarrow 0^+_1$ transition measured in $^{52}$Cr is around twice that of $^{50}$Ti \cite{Poves01}. In general, constraints on the decay intensities observed between the mainly non-yrast states and the {\it V=2} seniority states are valuable targets for experiments aiming to understand the microscopic behaviors in {\it A}$\approx$50 semi-magic nuclei, particularly the {\it B(E2; $0_2^+$\,$\rightarrow$\,$2_2^+$)} transition strength for the {\it $\nu$2p-2h} configuration. A deeper insight into the interplay between configurations will be obtained from a detailed comparison between calculations and experimental data.

In this article we report on the most sensitive study of the $\beta$ decay of $^{50}$Sc to $^{50}$Ti performed to date, using the GRIFFIN spectrometer at TRIUMF-ISAC \cite{Svensson2013,Rizwan2016,Garnsworthy2017A,Garnsworthy2018}.
The analysis of the excited states in $^{50}$Sc populated from the $\beta$ decay of the $^{50}$Ca beam in this work have been previously reported in Ref. \cite{Garnsworthy_2017B}. The new data for $^{50}$Ti presented here are compared to shell-model calculations utilizing both phenomenologically-derived interactions employed in the {\it pf} shell as well as an {\it ab initio} based interaction built in the valence-space in-medium similarity renormalization group framework.

\section{\label{sec:expsetup}Experimental details}
The isotope $^{50}$Ca ($T_{1/2}$ = 13.9(6)\,s \cite{Warburton1970}) was produced from reactions induced in a 22.49\,g/cm$^2$ Ta target by a 500\,MeV proton beam delivered by the TRIUMF Cyclotron \cite{Bylinskii2013}. The position of the 60\,$\mu$A proton beam on the ISOL target was continuously rastered such that a tighter proton beam spot could be used to induce a higher localized power density in the Ta material. The calcium atoms created in the target that diffused out of the material were ionized using resonant-laser ionization and accelerated to 20\,keV, mass separated and delivered to the experimental station. The typical beam intensity of $^{50}$Ca was $\sim 10^6$\,ions/s. Small amounts of surface-ionized $^{50}$K ($T_{1/2}$=472(4)~ms \cite{Langevin1983}) and $^{150}$Tb ($T_{1/2}$=3.5~hrs \cite{Basu2013,Haenni1977}) were also present in the beam.

The ions were stopped in a mylar tape at the central focus of the Gamma-Ray Infrastructure For Fundamental Investigations of Nuclei (GRIFFIN) spectrometer \cite{Svensson2013,Rizwan2016,Garnsworthy2017A,Garnsworthy2018}. GRIFFIN consists of an array of 16 high-purity Germanium (HPGe) clover detectors coupled to a series of ancillary detectors. Fifteen HPGe clovers were used in the present work. An array of plastic scintillator paddles (SCEPTAR) was used for the detection of $\beta$ particles. Four cerium-doped lanthanum bromide (LaBr$_3$(Ce)) scintillators were installed in the array but the data from them was not used in this work. The GRIFFIN clovers were positioned at a source-to-detector distance of 11~cm from the implantation point. A 20\,mm thick delrin plastic absorber shell was placed around the vacuum chamber to prevent $\beta$ particles from reaching the HPGe detectors while minimizing the flux of Bremsstrahlung photons created as the $\beta$ particles were brought to rest.

In order to study the longer-lived $^{50}$Sc daughter (T$_{1/2}$ = 102.5(5)\,s) activity, the beam was continuously delivered to the experimental station with the tape stationary. Data were collected in this way for a period of 5 hours. As previously reported \cite{Garnsworthy_2017B}, a series of short cycles were collected to clearly distinguish the activity of the $^{50}$Ca beam. In addition, a longer tape cycle of 1~min $^{50}$Ca implantation and 10~min decay was collected for a short time and used in the current analysis.

Energy and timing signals were collected from each detector using the GRIFFIN digital data acquisition system \cite{Garnsworthy2017A}, operated in a triggerless mode. HPGe energy and efficiency were calibrated using standard radioactive sources of $^{133}$Ba, $^{152}$Eu, $^{60}$Co and $^{56}$Co with the necessary corrections for coincidence summing applied.

\section{\label{sec:results}Data analysis and results}

\subsection{\label{ssec:gammaspec}Gamma-ray energy spectra}

Events observed by individual GRIFFIN clover detectors were time-correlated to produce $\gamma$-ray addback spectra and provided the principal tool for offline analysis of $^{50}$Sc decay. Time-correlated $\gamma$-ray addback hits were used to construct $\gamma$-$\gamma$ matrices in order to establish branching ratios and coincidence relationships in $^{50}$Ti, with the option of requiring coincidences with $\beta$ electrons detected in SCEPTAR.

\begin{figure}[ht]
		\includegraphics[angle=0,width=0.95\linewidth]{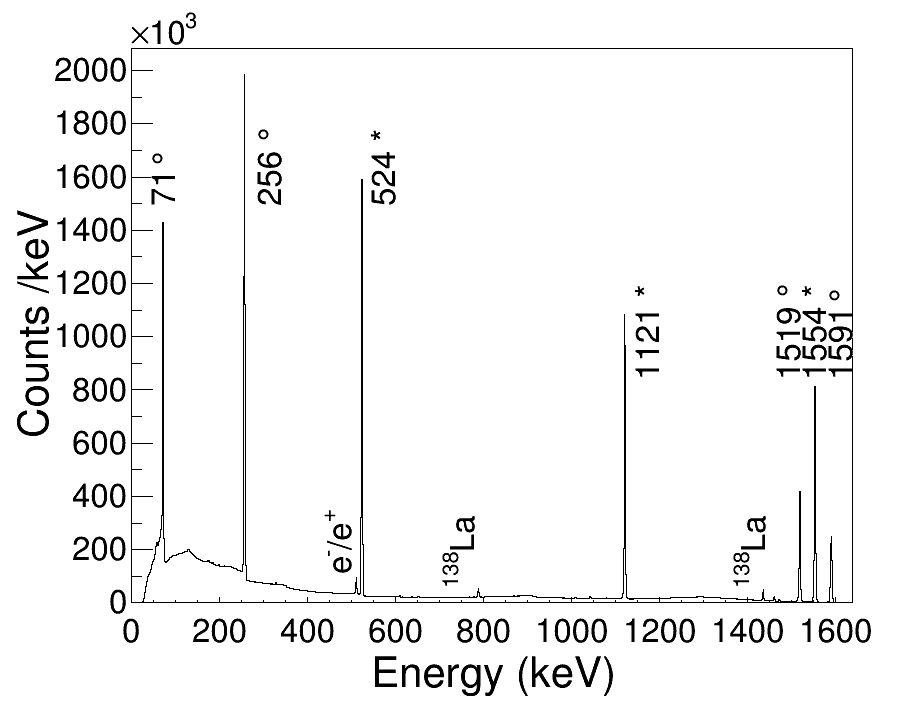}
		\centering
        \caption{$\gamma$-ray addback energy spectrum collected with GRIFFIN following the decay of $^{50}$Ca. Internal transitions belonging to $^{50}$Sc ($^{\circ}$) and $^{50}$Ti ($^{*}$) are indicated. Room background lines are also labelled, including those from the $^{138}$La natural radioactivity present in the LaBr$_3$(Ce) detectors.}
        \label{fig_energy}
		\end{figure}

An addback energy spectrum is shown in Figure \ref{fig_energy} for $\gamma$-ray energies below 1.6\,MeV, encompassing the most intense transitions observed following the decay of $^{50}$Ca and $^{50}$Sc. The intensity of the 1554\,keV 2$^{+}$ $\rightarrow$ 0$^{+}$ transition in $^{50}$Ti yields a total of $\approx$ 2.8~$\times$10$^{9}$ $^{50}$Sc decays. The fraction of contaminant nuclei in the beam was assessed using $\gamma$ rays emitted following $\beta$-decay and $\beta$-delayed neutron emission of $^{50}$K and the EC decay of $^{150}$Tb, transmitted to GRIFFIN as $Q=1^{+}$ and 3$^{+}$ charge states, respectively. The total contaminant activity detected in the chamber was $\approx$ 1 $\%$ relative to the 1554\,keV transition.

    \begin{figure}[ht]
		\includegraphics[angle=0,width=0.95\linewidth]{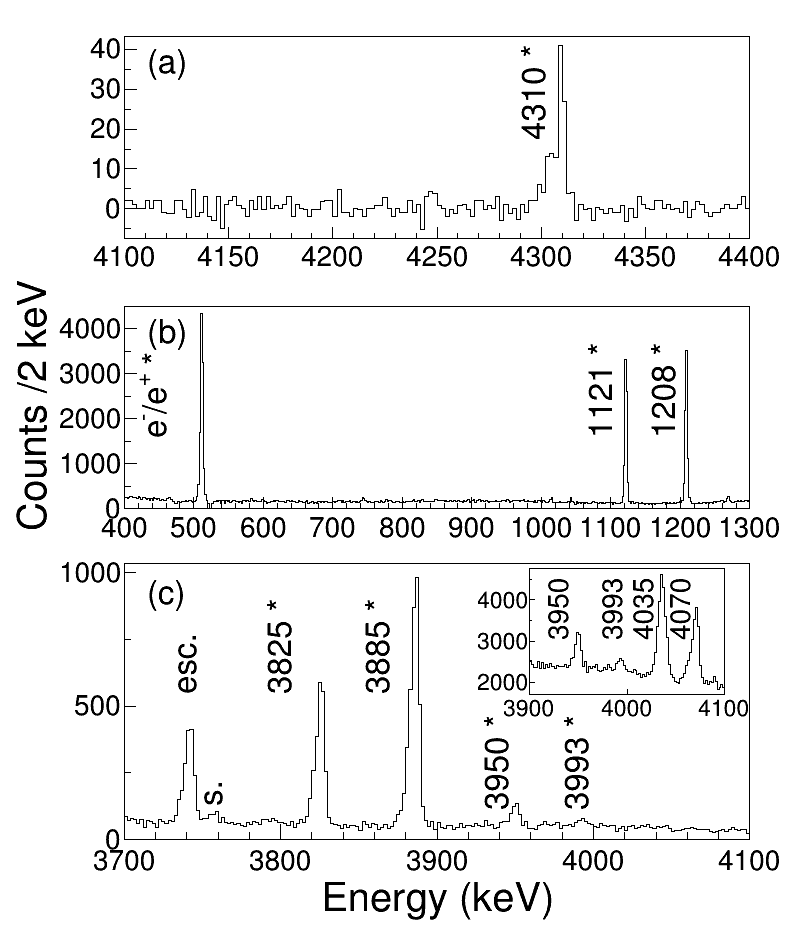}
		\centering
        \caption{$\gamma$-ray gated addback energy spectra showing coincidence gates with the (a) 1130\,keV transition feeding the 4310\,keV $2^+$ state, (b) 2618\,keV $\gamma$ ray and the (c) 1554\,keV transition. A $\gamma$-ray singles energy spectrum comprised of the sum of all addback hits is shown in the inset of panel (c) for a subset of energies, where additional transitions at 4035 and 4070\,keV are observed following the $\beta$ and $\beta$-n decay of $^{50}$K, respectively. Peaks labelled with an asterisk are transitions identified in $^{50}$Ti. Summing (\textit{s.}) and escape (\textit{esc.}) peaks are also indicated.}
        \label{fig_energy2}
		\end{figure}

\subsection{\label{ssec:levelscheme}Level scheme}

\begin{figure*}[ht]
		\centering
		\includegraphics[angle=0,width=0.95\linewidth]{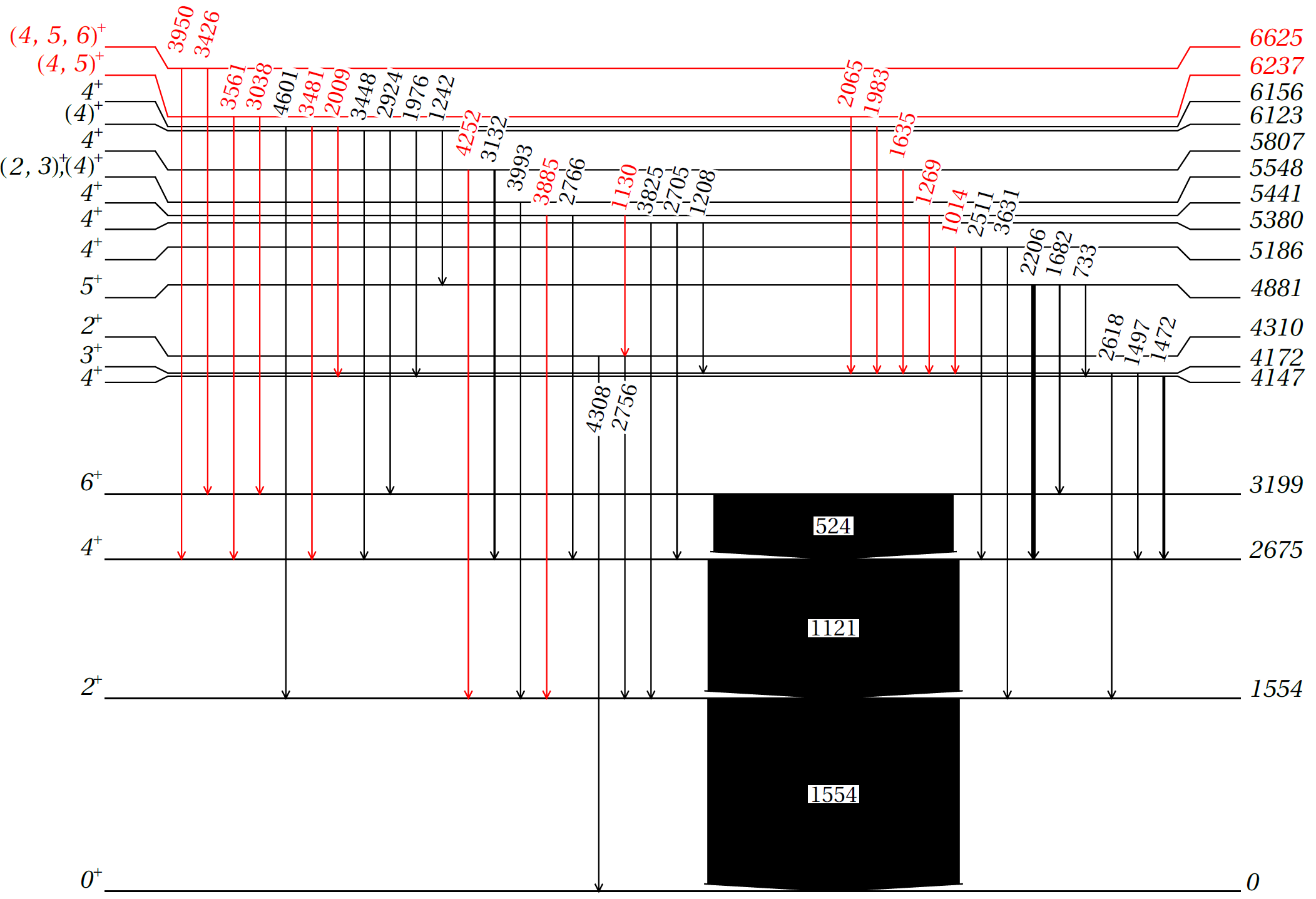}
        \caption{Level scheme of the levels in $^{50}$Ti constructed following the $\beta$ decay of $^{50}$Sc. Line width is proportional to the observed $\gamma$-ray intensity. Black arrows correspond to previously observed $\gamma$-ray transitions, while red indicates those observed in the current work and have been assigned to the level scheme of $^{50}$Ti for the first time. 
        }
        \label{fig_lvl_scheme}
		\end{figure*}

The level scheme observed in the decay of $^{50}$Sc to $^{50}$Ti was constructed on the basis of a $\gamma - \gamma$ coincidence analysis. Examples of $\gamma$-ray gated addback energy spectra are shown in Figure \ref{fig_energy2}. Figure \ref{fig_lvl_scheme} shows the placement of $\gamma$ rays observed in the current work into the $^{50}$Ti level scheme. The width of the arrows indicates the intensity relative to that of the 1554\,keV transition. In addition to the 1554\,keV decay from the first excited state, only one other transition is observed to decay directly to the ground state, de-exciting the $J^\pi$=$2_2^+$ state at 4310\,keV (Figure \ref{fig_energy2} (a) gated on the 1130\,keV feeding transition). Otherwise all excited states eventually feed the yrast, $J^+$=$2^{+}$,$4^{+}$ and $6^{+}$ states. For this reason, gates placed upon the $\gamma$ ray of interest (i.e. gating from above) were often the most useful concerning placement in the level scheme. Transitions were placed according to their observed coincidence (or non-coincidence) with the strongest transitions in $^{50}$Ti, notably the 524, 1121 and 1554\,keV $\gamma$ rays, and via comparison of $\gamma$-ray energies with the energy difference $\Delta E$ between previously known excited states. Feeding transitions were observed in a few cases above the 3199\,keV $6^+$ state (at 4147, 4172 and 4310\,keV) helping to constrain the measurement of direct $\beta$ decay branching to these states.

The 3132\,keV $\gamma$ ray de-exciting the 5807\,keV state to the 2675\,keV $4^+$ state is observed to interfere with a different transition (the 2618\,keV transition de-exciting the yrast $3^+$ state at 4172\,keV) through an energy-coincidence of the associated single-escape peak at 2621\,keV. This was confirmed by gating on the 2618\,keV transition (Figure \ref{fig_energy2}b) where strong coincidences are observed with both a 511\,keV escape photon and the 1121\,keV transition. The intensity of the 2618\,keV $\gamma$ ray has been corrected for this contribution (Section \ref{ssec:gintensities}).

There is good agreement between the current work and that from a previous $\beta$-decay study reported by Alburger {\it et al.} \cite{Alburger1984} as well as with Ruyl {\it et al.} \cite{Ruyl84} who utilized thermal neutron capture on metallic titanium targets. In the present work, the level scheme has been expanded with a number of levels and transitions observed for the first time in $\beta$ decay. This is discussed in detail in the following Sections.

\subsection{\label{ssec:gintensities}Gamma-ray relative intensities}

Intensities of $\gamma$ rays observed following the decay of $^{50}$Sc are given in Table \ref{table:intensities}. Branching ratios of $\gamma$ rays from each excited state are also provided. $\gamma - \gamma$ coincidence requirements have been applied where possible in order to isolate the transitions of interest and obtain the optimum peak-to-background ratio for extracting relative intensities. This is especially important for weaker branches obscured by the Compton-scattered background arising from more intense transitions (mainly the 1121 and 1554\,keV $\gamma$ rays). A procedure similar to the `gating from below' method employed by Kulp {\it et al.} \cite{Kulp07} has been used here to obtain the normalized intensity $I_\gamma$ for the transitions of interest. Modifications to the overall detection efficiency due to coincidence timing restrictions and the angular coverage of GRIFFIN compared to singles data are assumed to be $\approx$ 1.0 (see for example, Ref. \cite{Garrett12}). The gating transition was chosen to de-excite states below that of the transition of interest with the 524, 1121 and 1554\, keV $\gamma$ rays being the most common choice due to their well-characterized branching ratios. The transition of interest directly feeds the level depopulated by the gating transition in all cases.

All intensities provided in Table \ref{table:intensities} have been corrected for summing effects using an empirical method described in Ref. \cite{Garnsworthy2018}. In cases where transition intensities are extracted from $\gamma$-ray addback hits or $\gamma$-ray single hits without any coincidence conditions applied (i.e. the 524, 1121, 1554, 3427, 3950, 3993 and 4601\,keV transitions and the 2756 and 4308\,keV transitions respectively), summing corrections are obtained using a $\gamma - \gamma$ coincidence matrix with the requirement that HPGe clovers be located at 180~$^\circ$ with respect to each other. A normalization factor is included to reconcile the difference in combinatorial efficiency between the intensity obtained in singles and the summing correction. This may arise due to asymmetries in the detector array where the availability of single crystals (clovers) differs from the number of crystal (clover) pairs at 180~$^\circ$. A similar method is employed to extract summing corrections for the remaining transitions in Table \ref{table:intensities} extracted from coincidence measurements. In the coincidence case the summing correction factors are specific to the transition of interest as well as the choice of the gating transition. Much care must be taken when constructing the necessary coincidence matrices used to determine these factors in order that the same experimental conditions are applied to them as to the experimental data.

Data in Table \ref{table:intensities} are compared to previously reported measurements where available. A total of 38 transitions have been identified in $^{50}$Ti. The vast majority of the $\gamma$-ray intensity ($\approx$ 99~$\%$) is contained in the 524, 1121 and 1554\,keV transitions. The remaining $\approx$ 1~$\%$ includes many weak transitions, 25 of which are observed here in $\beta ^{-}$ decay for the first time, with around half of these transitions having not been reported in any previous experiment. Relative intensities are determined to below 10$^{-3}$ $\%$ that of the 1554\,keV transition to the ground state, which is a factor of $\approx$15 lower than that of the 3825\,keV transition identified by Alburger {\it et al} \cite{Alburger1984}.

\begin{table*}[ht]
    \caption{Energies, intensities and branching ratios of $\gamma$ rays measured in the beta decay of $^{50}$Sc. $I_{\gamma}$ is expressed relative to the 1554\,keV transition to the ground state of $^{50}$Ti determined using a $\gamma - \gamma$ coincidence matrix except where noted. Data are compared to previous measurements where available. See text for details.}
  	\label{table:intensities}
	\begin{ruledtabular}
    \begin{tabular}{c|ccccc|cc}
    & \multicolumn{5}{c|}{This work} & \multicolumn{2}{c}{Previous Works} \\
			$E_i$\footnote{Ref. \cite{Chen2019} except for 6237 and 6625\,keV states which are determined using the (weighted) average energies of depopulating transitions, corrected for nuclear recoil.} (keV)  & $J_i^{\pi}$\footnote{Ref.\cite{Chen2019} or current work.} & $E_{\gamma}$\footnote{Systematic and statistical uncertainties added in quadrature.} (keV) & $E_f$ (keV) & BR & $I_\gamma$ & BR & $I_\gamma$ \\
	&&&&&&Ref.\cite{Chen2019}&Ref.\cite{Alburger1984} \\
			\hline
1553.8 &2$^+$& 1553.8(2)\footnote{$I_{\gamma}$ obtained in singles mode using $\gamma - $ray addback energy spectrum.}&0.0& 100 & 100 & 100 & 100 \\
2674.9 &4$^+$& 1121.0(2)\footnotemark[4]&1553.8& 100 & 100.1~\textit{22} & 100  & 99.54~\textit{90} \\
3198.7 &6$^+$& 523.7(2)\footnotemark[4]&2674.9& 100 & 88.5~\textit{17} & 100  & 88.74~\textit{150} \\
4147.2 &4$^+$& 1472.3(2)\footnote{$I_{\gamma}$ from coincidence gate placed at 1121\,keV.} &2674.9& 100 & 0.630~\textit{14} & 100  & 0.61~\textit{4} \\
4171.9 &3$^+$	& 1497.1(2)\footnotemark[5] &2674.9& 41.1~\textit{31} & 0.0365~\textit{26} & 48~\textit{3}  & \textless0.10 \\
		&& 2618.0(2)\footnote{$I_{\gamma}$ from coincidence gate placed at 1554\,keV.} &1553.8& 100.0~\textit{28} & 0.0889~\textit{28} & 100~\textit{6}  & \textless0.30 \\
4310.0 &2$^+$	& 2756.0(2)\footnote{$I_{\gamma}$ obtained in singles mode using $\gamma - $ray single-hit energy spectrum.} &1553.8& 100.0~\textit{60} & 0.00381~\textit{24} & 100~\textit{10}  & \\
		&& 4307.8(11)\footnotemark[7]&0.0& 26.8~\textit{35} & 0.00102~\textit{12} & 19.6~\textit{21}  & \\
4880.6 &5$^+$	& 733.4(2)\footnote{$I_{\gamma}$ from coincidence gate placed at 1472\,keV.} &4147.2& 2.14~\textit{6} & 0.02683~\textit{70} & 2.12~\textit{20}  & \\
		&& 1682.0(2)\footnote{$I_{\gamma}$ from coincidence gate placed at 524\,keV.}&3198.7& 16.96~\textit{44} & 0.2131~\textit{52} & 8.3~\textit{24}  & 0.28~\textit{3} \\
		&& 2205.8(2)\footnotemark[5] &2674.9& 100.0~\textit{18} & 1.256~\textit{30} & 100~\textit{6}  & 1.27~\textit{3} \\
5186.1 &4$^+$	& 1014.6(8)\footnote{$I_{\gamma}$ from coincidence gate placed at 2618\,keV.} &4171.9& 2.24~\textit{37} & 0.00152~\textit{25} &  & \\
		&& 2511.3(2)\footnotemark[5] &2674.9& 100.0~\textit{21} & 0.0678~\textit{18} & 100~\textit{7}  & \\
		&& 3631.5(10)\footnotemark[6] &1553.8& 38.2~\textit{13} & 0.02588~\textit{83} & 40.4~\textit{24}  & \\
5379.9 &4$^+$	& 1207.7(2)\footnotemark[10] &4171.9& 49.5~\textit{31} & 0.0531~\textit{33} & 54.8~\textit{31}  & \textless0.10 \\
		&& 2705.1(2)\footnotemark[5] &2674.9& 100.0~\textit{19} & 0.1071~\textit{27} & 100~\textit{7}  & 0.105~\textit{16} \\ 
		&& 3825.1(10)\footnotemark[6] &1553.8& 11.28~\textit{58} & 0.01209~\textit{61} & 12.8~\textit{10} & 0.044~\textit{10} \\
5440.7 &4$^+$	& 1130.4(3)\footnote{$I_{\gamma}$ from coincidence gate placed at 2756\,keV.} &4310.0& 2.13~\textit{23} & 0.00245~\textit{25} &  &\\
		&& 1268.9(8)\footnotemark[10] &4171.9& 1.31~\textit{24} & 0.00152~\textit{28} &  &\\
		&& 2765.7(2)\footnotemark[5] &2674.9& 100.0~\textit{42} & 0.1152~\textit{51} & 100  & 0.145~\textit{18}\\
		&& 3885.5(10)\footnotemark[6] &1553.8& 18.67~\textit{98} & 0.02151~\textit{77} &  &\\
5547.8 &(2, 3)$^+$, (4)$^+$\footnote{Spin and parity assignments based upon $\beta$ feeding from the $^{50m}$Sc and $^{50}$Sc parent states, respectively. See Table \ref{table:BetaBR_LogFT}.}	& 2872.8\footnote{Not observed. 1.22$\times I_{\gamma}^{3993\,keV}$ from Ref.\cite{Ruyl84}.}&2674.9& 100.0~\textit{168}& 0.00075~\textit{13} & 100~\textit{6} &\\
				&& 3993.2(11)\footnotemark[4]&1553.8& 81.7~\textit{63} & 0.00061~\textit{9} & 82~\textit{5}  &\\
5806.5 &4$^+$	& 1635.2(8)\footnotemark[10] &4171.9& 0.91~\textit{9} & 0.00249~\textit{25} &  &\\
		&& 3132.2(2)\footnotemark[5] &2674.9& 100.0~\textit{22} & 0.2747~\textit{75} & 100  & 0.251~\textit{15}\\
		&& 4251.8(10)\footnotemark[6] &1553.8& 12.3~\textit{12} & 0.0338~\textit{32} &  &\\
6123.1 &(4)$^+$	& 1242.4(2)\footnote{$I_{\gamma}$ from coincidence gate placed at 2206\,keV.} &4880.6& 100.0~\textit{35} & 0.0512~\textit{20} & 100~\textit{7}  & \\
		&& 1975.6(2)\footnotemark[8] &4147.2& 24.6~\textit{11} & 0.01262~\textit{42} & \textless15  & \\
		&& 2924.3(2)\footnotemark[9] &3198.7& 31.5~\textit{16} & 0.01615~\textit{67} & 31~\textit{7}  & \\
		&& 3448.4(2)\footnotemark[5] &2674.9& 24.8~\textit{14} & 0.01272~\textit{57} & 20~\textit{5}  & \\
6156.4 &4$^+$	& 1983.4(8)\footnotemark[10] &4171.9& 4.88~\textit{103} & 0.00044~\textit{9} &  & \\	
		&& 2008.9(2)\footnotemark[8] &4147.2& 100.0~\textit{34} & 0.00903~\textit{34} &  & \\
		&& 3481.4(3)\footnotemark[5] &2674.9& 58.5~\textit{41} & 0.00528~\textit{34} &  & \\
		&& 4600.6(12)\footnotemark[4]&1553.8& 12.6~\textit{13} & 0.00114~\textit{11} & 100  & \\
6236.9(2) &(4,5)$^+$	& 2064.8(8)\footnotemark[10]&4171.9& 2.97~\textit{44} & 0.00066~\textit{10} &  &\\
			&& 3038.1(2)\footnotemark[9] &3198.7& 61.1~\textit{27} & 0.01355~\textit{52} &  &\\
			&& 3560.9(10)\footnotemark[5] &2674.9& 100.0~\textit{27} & 0.02217~\textit{69} &  &\\
6625.4(3) &(4,5,6)$^+$	& 3426.6(3)\footnotemark[4]&3198.7& 63.7~\textit{86} & 0.00132~\textit{15} &  &\\
			&& 3950.1(10)\footnotemark[4]&2674.9& 100.0~\textit{72} & 0.00208~\textit{15} &  &\\
    \end{tabular}
    \end{ruledtabular}
\end{table*}

\subsection{\label{sec:bfeeding}Beta-decay branching ratios}

$\beta$-decay branching ratios to excited states in $^{50}$Ti were determined based on the coincidence relationships established in the current work and the observed intensities of $\gamma$ rays with the appropriate corrections for internal conversion. The $\beta$-decay branching ratios from this work are reported in Table \ref{table:BetaBR_LogFT}. Conversion coefficients were calculated using the {\it BrIcc-FO} (frozen orbitals) formalism \cite{Kibedi2008} with transition multipolarities inferred from the spin-parities of the initial and final states summarized in Table \ref{table:BetaBR_LogFT}. While mixing ratios compiled in Ref. \cite{Chen2019} are used where available, the leading order multipolarity is assumed to dominate for most mixed transitions. It should be noted however that the corrections are small: of the order $10^{-4} - 10^{-5}$ in $^{50}$Ti.

\begin{table*}[ht]
    \caption{$\beta$-branching ratios and calculated log({\it ft}) values observed in the $\beta$ decay of $^{50}$Sc as compared to previously reported values. Spin and parities are assigned in the current work by comparison with both existing assignments and with the typical values for log({\it ft}) compiled in Ref. \cite{Singh1998}. In some cases further restrictions have been imposed upon spin and parity assignments by considering the initial and final states in $^{50}$Ti connected by internal $\gamma$ ray transitions (see text for details). The excited states at 6237 and 6625\,keV are reported here for the first time.}
  	\label{table:BetaBR_LogFT}
	\begin{ruledtabular}
    \begin{tabular}{cccccccc}
           & This work & & & & Literature \cite{Chen2019} & \\
			$E_i$ (keV) & $I_{\beta}$\footnote{$\beta$ branch $\times$ 100\,$\%$. For intensity per 100 decays of the $2_1^+$ level, multiply by 1.00.} (\%) & log{\it ft} & $B$(GT)\footnote{$B$(GT) = (6144.5/{\it ft})\,$\times$\,10$^3$. $B$(GT) $<$ 1.5$\times 10^{-5}$ for transitions from the $^{50}$Sc ground state to the 1554 and 4310\,keV states.}\,(10$^{-3}$) & $J_{f}^{\pi}$ & $I_{\beta}$ \footnote{Intensity per 100 decays.} & log{\it ft} & $J_{f}^{\pi}$ \\
            \hline
\\

\multicolumn{8}{l}{\bf{Decay from $^{50}$Sc ground state, $J^\pi=5^+$}}\\

1554 & $<$1.8 & $>$11.6 & &$2^+$ & $<$1.3 \footnotemark[4] & $>$11.76 \footnote{Ref. \cite{Alburger1984}. This {\it $\Delta$J}=3 $\beta$ transition is erroneously reported as having a log$ft$ value of $>7.9$ in Ref. \cite{Elekes2011}.} & $2^+$ \\
2675 & 9.0~\textit{17} & 6.64~\textit{9} &1.41~\textit{32}& $4^+$ & 8.4~\textit{18} & 6.7~\textit{1} & $4^+$ \\
3199 & 88.3~\textit{17} & 5.39~\textit{1} &24.92~\textit{58}& $6^+$ & 88.4~\textit{15} & 5.39~\textit{1} & $6^+$ \\
4147 & 0.582~\textit{14} & 7.00~\textit{2} &0.614~\textit{24}& $4^+$ & 0.61~\textit{4} & 6.98~\textit{3} & $4^+$ \\
4172 \footnote{Doublet comprised of 4171.968(17)\,keV $3^+$ state and 4172.5(4)\,keV $(2)^+$ state \cite{Chen2019}. Only the $3^+$ state appears to be fed (see Sec. \ref{ssec:lev_4172}).} & $<$0.0657~\textit{49} & $>$7.9 & &$3^+$ & $<$0.35 & $>$7.2 & $3^+$ \\
4310 & $<$0.00257~\textit{37} & $>$11.9 && $2^+$ & & & $2^+$ \\
4881 & 1.445~\textit{32} & 6.03~\textit{2} 	&5.73~\textit{26}& $5^+$ & 1.55~\textit{5} & 6.00~\textit{2} & $5^+$ \\
5186 & 0.0953~\textit{22} & 6.91~\textit{2} 	&0.756~\textit{36}& $4^+$ & & & $(3,4)^+$ \\
5380 & 0.1724~\textit{47} & 6.44~\textit{2} 	&2.21~\textit{11}& $4^+$ & 0.149~\textit{19} & 6.50~\textit{6} & $4^+$ \\
5441 & 0.1408~\textit{53} & 6.46~\textit{2} 	&2.13~\textit{10}& $4^+$ & 0.145~\textit{18} & 6.45~\textit{6} & $4,5^+$ \\
5548 & 0.00137~\textit{16} & 8.33~\textit{6} 	&0.0287~\textit{43}& $(4)^+$ & & & $(4^+)$ \\
5807 & 0.3112~\textit{84} & 5.61~\textit{3} 	&15.1~\textit{11}& $4^+$ & 0.251~\textit{15} & 5.71~\textit{4} & $4^+,5^+$ \\
6123 & 0.0927~\textit{25} & 5.56~\textit{4} 	&16.9~\textit{16}& $(4)^+$& & & $(4^+)$ \\
6156 & 0.01590~\textit{52} & 6.25~\textit{4} 	&3.46~\textit{33}& $4^+$ & & & $(2,3,4)^+$ \\
6237 & 0.03638~\textit{96} & 5.71~\textit{4} 	&12.0~\textit{11}& $(4,5)^+$ & & & \\
6625 & 0.00340~\textit{22} & 5.35~\textit{9} 	&27.4~\textit{63}& $(4,5,6)^+$ & & & \\
\\
\multicolumn{8}{l}{{\bf Decay from $^{50m}$Sc isomeric state, $J^\pi=2^+$}}\\
1554 & $\leq$0.1 & $\geq$6.7 &$<$1.2 &$2^+$ &  &  & $2^+$ \\
4172 \footnotemark[5] & $\leq$0.0657~\textit{49} & $\geq$5.63~\textit{6} &$<$14.4& $3^+$ & $<$0.35 & $>$7.2 & $3^+$, $(2)^+$ \\
4310 & $\leq$0.00257~\textit{37} & $\geq$6.95~\textit{9} &$<$0.69& $2^+$ & & & $2^+$ \\
5548 \footnote{No confirmed transition to the 3198.7\,keV $6^+$ state (See Sec. \ref{ssec:lev_5548}).} & 0.00137~\textit{16} & 6.17~\textit{8} &$<$4.1& $(2,3)^+$ & & & $(4^+)$ \\
    \end{tabular}
    \end{ruledtabular}
\end{table*}

The ground state spin of $^{50}$Sc is established as $J^\pi=5^+$ and therefore a direct $\beta$ decay to the $0^+$ ground state of $^{50}$Ti would be a fourth forbidden unique transition. We therefore make the assumption that there is negligible direct feeding to the $0^+$ ground state, as was also done by Alburger {\it et al.} \cite{Alburger1984}.

An upper limit for the $\beta$ decay branch from the 257\,keV isomeric state of $^{50}$Sc feeding the 1554\,keV $2^+$ state in $^{50}$Ti was reported by Alburger {\it et al.} \cite{Alburger1984}. In our previous reporting from the present work, the upper limit for this decay branch was reduced from $<$2.5\% to $<$1\% \cite{Garnsworthy_2017B} by observation of the number of 1554\,keV $\gamma$ rays collected in singles relative to those in coincidence with the 1121\,keV transition, extracted from short-decay cycles. Direct $\beta$ decay from the $2^+$ isomeric state in $^{50}$Sc has been observed unequivocally in the current work through a detailed examination of the $\beta$ feeding and log($ft$) values. 
Evidence of $\beta$ feeding from the isomer is found in the case of the 4172\,keV $3^+$ state and the 4310\,keV $2^+$ state and are detailed in Table \ref{table:BetaBR_LogFT}. The isomer may also feed the 5548\,keV state given a possible spin and parity assignment of (2,3)$^+$. The branching ratios to these states are reported as upper limits because of possible unobserved $\gamma$-ray transitions feeding these levels.

Following further analysis of the $^{50}$Ca daughter activity observed in the short-decay cycles, the $\beta$-decay branching ratio of the 257\,keV isomeric state in $^{50}$Sc towards the first 2$^+$ state in $^{50}$Ti is reduced from $<$1\% to $<$0.1\% at the 95\% confidence level. The short-decay cycle comprised a 2\,s tape move and background measurement followed by a 3\,s $^{50}$Ca beam implantation period and 3\,s of decay. The ratio of the number of counts in the 1554\,keV peak acquired in single-hit mode compared to coincidence mode was measured around 2\,s after the beginning of the decay period (around 7\,s into the full cycle). The experimental ratio was compared to the calculated ratio of daughter activities resulting in (i) the detection of 1554\,keV $\gamma$ rays (i.e. all activities) and (ii) the detection of 1554\,keV $\gamma$ rays in coincidence with the 1121\,keV transition and decays from higher-lying levels. The ratio of activities during the cycle can be calculated using the half lives of the parent and daughter nuclei and by assuming values for the $\beta$-branching ratio of the 257\,keV isomer $BR_{\beta} (^{50m}Sc)$ and the fraction of decays $\delta (^{50}Ti)$ from the $^{50}$Sc ground state that proceed via excited states in $^{50}$Ti, excluding the 2$_1^+$ state. The calculated activity ratio is compared to the experimental ratio which constrains the range of possible values for $BR_{\beta} (^{50m}Sc)$ and $\delta (^{50}Ti)$ within uncertainty limits. A multivariate analysis of $BR_{\beta} (^{50m}Sc)$ and $\delta (^{50}Ti)$ shows that the $\beta$ feeding of the $2_1^+$ state by the $^{50}$Sc ground state is $\approx$\,0 if $BR_{\beta}(^{50m}Sc)$ $<$0.1\%, within 2 standard deviations. These observations are also consistent with the upper limit for the total $\beta$ feeding to the first $2_1^+$ state in $^{50}$Ti of $<$1.8\% in Table \ref{table:BetaBR_LogFT} ($<$4\% at the 95\% confidence level) obtained using the continuous-implantation mode which has a reduced sensitivity to the feeding from the short-lived isomeric state.

The $I_{\beta}$ values in Table \ref{table:BetaBR_LogFT} have been normalised with respect to the sum of the 1554 and 4308\,keV transition intensities feeding the $^{50}$Ti ground state.


\subsection{\label{ssec:logft}log({\it ft}) values and $B(GT)$ transition strength}

\begin{figure}[ht]
	\includegraphics[angle=0,width=0.95\linewidth]{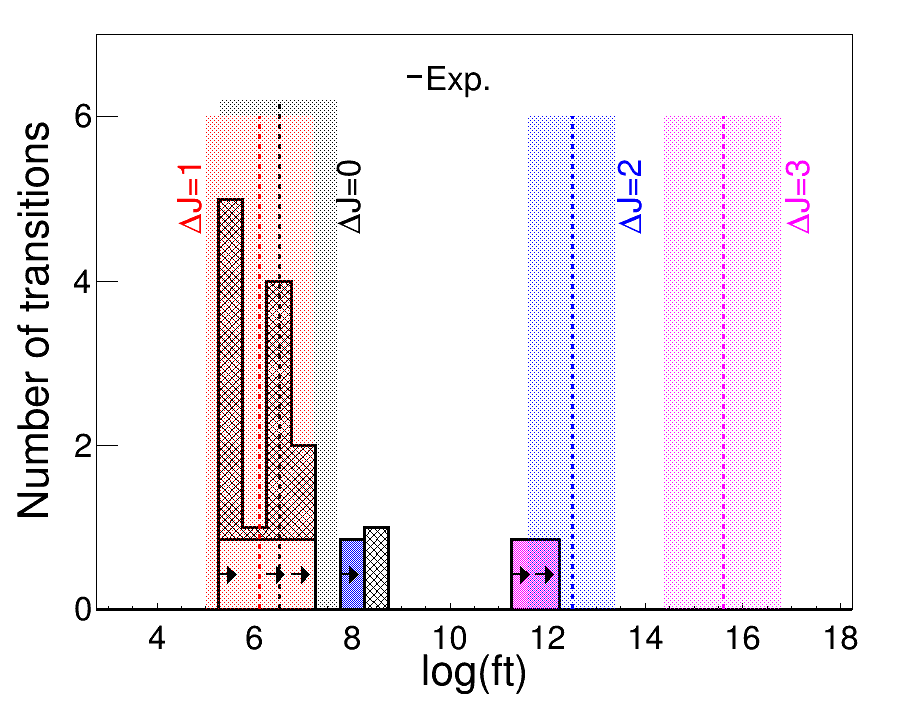}
	\centering
    \caption{Histograms summarizing log({\it ft}) values for $^{50}$Sc $\beta$ decay determined from experimental branching ratios. The hatched histogram indicates $\beta$ transitions from the {\it 5$^+$} ground state whereas the open-area histogram (scaled for clarity) corresponds to transitions from both the ground state and the 257\,keV {\it 2$^+$} isomeric state. The open-area histogram corresponds to {\it $\Delta$J}=0,1 transitions except where shaded blue ({\it $\Delta$J}=2) or magenta ({\it $\Delta$J}=3). Lower-limits associated with individual bins are indicated by arrows. Data are compared to typical log({\it ft}) values compiled in Ref. \cite{Singh1998} for different {\it $\Delta$J} ({\it $\Delta\pi$=no}). See text for details.}
    \label{fig:LogFT}
\end{figure}

Table \ref{table:BetaBR_LogFT} lists the log({\it ft}) values obtained using the log({\it ft}) calculator of Ref. \cite{logft_nndc} for the decay of the $5^+$ ground state and $2^+$ isomeric state in $^{50}$Sc. The $\beta$-branching ratios determined in the current work are used as inputs to the calculation in addition to excitation energies of states populated in $^{50}$Ti, the half-lives of the $5^+$ ground state and $2^+$ isomeric state in $^{50}$Sc and the $\beta$-decay {\it Q} value of 6.884(15)\,MeV \cite{ame2016}. 
The log({\it ft}) values from this work are plotted in Figure \ref{fig:LogFT} in comparison to the range of typical values for $\beta^{-,+}$ and $EC$ decaying nuclei compiled by Singh {\it et al.} \cite{Singh1998}. 

A good agreement is found for log({\it ft}) values obtained in the current work compared to those included in the 2019 nuclear data evaluation by Chen and Singh \cite{Chen2019} and support the assigned spin and parities given in column 4 of Table \ref{table:BetaBR_LogFT}. As is evident in Figure \ref{fig:LogFT}, $^{50}$Sc $5^+$ ground state $\beta$ decay to known $4^+, 5^+$ and $6^+$ levels proceeds with an experimental weighted-mean log($ft$) of around 6.3, consistent with a dominance of {\it L=0} allowed Fermi and Gamow-Teller $\beta$ transitions. The $^{50m}$Sc $2^+$ isomeric state $\beta$ decay to known $2^+$ and $3^+$ levels also have log($ft$) values consistent with allowed $\Delta J=0,1$ transitions.

Gamow-Teller transition strengths have been calculated for $\Delta J=0,1$ $\beta$ transitions from the $^{50}$Sc ground state and $^{50m}$Sc isomeric state to levels in $^{50}$Ti. The values are shown in Table \ref{table:BetaBR_LogFT}. The results compare favourably with those originally reported by Alburger et al. for the 2675, 3199, 4147, 4881, 5380, 5440 and 5807\, keV levels (Table 3 in Ref. \cite{Alburger1984}). The average superallowed $\addbar{\mathcal{F}t}$ value adopted in Table \ref{table:BetaBR_LogFT} is that reported by Hardy and Towner in their most recent review \cite{Hardy2020}. In the current work, the $B(GT)$ transition strength to levels in $^{50}$Ti is increased by more than a factor of 2 compared with Ref. \cite{Alburger1984}, mainly due to the 6123, 6156, 6237 and 6625\,keV levels where no $\beta$-feeding intensities were previously available.

\subsubsection{Previously known excited states}

A number of $\gamma$ rays are observed in the current work to de-excite levels previously identified by Alburger {\it et al.} \cite{Alburger1984} (and references therein) but were not originally placed in the $^{50}$Ti level scheme determined from $^{50}$Sc $\beta ^-$ decay. This includes excited states in Table \ref{table:BetaBR_LogFT} with energies 4881, 5441 and 5807\,keV where at least one transition was previously assigned depopulating each level.

For example, a transition observed in the current work with energy 733\,keV is assigned de-exciting the 4881\,keV state while the 1130, 1269 and 3885\,keV transitions are assigned to the 5441\,keV level. In addition, the 1635 and 4252\,keV transitions are assigned de-exciting the 5807\,keV state. In each of these cases the additional decay intensity introduced in the current work does not dramatically alter the previously established log({\it ft}) values. In general, the branching ratios of $\gamma$-ray transitions depopulating previously-observed levels in $^{50}$Ti show a good agreement with results from neutron-capture measurements \cite{Ruyl84}, although the 1682\,keV transition from the 4881\,keV level is around a factor of 2 larger than previously observed. A similar observation was made by Chen and Singh \cite{Chen2019} in their comparison of the results from the previous $\beta$-decay study \cite{Alburger1984} with the neutron-capture data. Alburger et al. observed an intensity of $\approx$\,22\% for the 1682\,keV transition relative to the 2206\,keV $\gamma$-ray.

In cases where no additional transitions are placed de-exciting a level (for example the $4^+$ states at 4147 and 5380\,keV) our log({\it ft}) values are similar to those of Ref. \cite{Chen2019} which incorporates the work of Alburger {\it et al.} \cite{Alburger1984}. The relative intensities of $\gamma$ rays de-exciting the 5380\,keV state show good agreement with that of Ref. \cite{Chen2019} although a significantly larger intensity for the 3825\,keV transition (around 40~$\%$ of the 2705\,keV transition intensity compared with 10~$\%$ in the current work) was reported in Ref. \cite{Alburger1984}. Consequently a weaker $\beta$ branch of 0.17~$\%$ is determined in the current work for the 5380\,keV state compared with Ref. \cite{Alburger1984} (0.22~$\%$) but is in agreement with the value included in Table \ref{table:BetaBR_LogFT} for the most recent evaluation \cite{Chen2019}.


Additional information regarding the spin and parity assignments of some previously observed excited states may be obtained through the placement of internal $\gamma$ ray transitions in $^{50}$Ti. For example the 5441 and 5807\,keV states have been assigned {\it J$^{\pi}$=4$^+$} in Table \ref{table:BetaBR_LogFT} as each state depopulates via transitions that feed the 1554\,keV $2_1^+$ state.

\section{\label{sec:discussion}Discussion}

In this work many excited states are observed, several of which have been observed in $\beta$ decay for the first time. These states require additional comment.

\subsection{\label{ssec:lev_4172}The 4172\,keV doublet}

It should be noted that the 4171.9\,keV $3^+$ state is an energy doublet appearing in close proximity to a 4172.5\,keV state with a tentative spin-parity assignment of $(2)^+$ \cite{Chen2019}, the latter decaying to the 1554\,keV $2_{1}^{+}$ via a 2618\,kev $\gamma$ ray. The $3^{+}$ state also decays to the 1554\,keV $2_{1}^{+}$ state via a 2618\,kev $\gamma$ ray and is assigned an additional 1497\,keV decay pathway to the 2675\,keV $4_{1}^{+}$ state. Both the 1497 and 2618\,keV transitions are observed here with intensities that are in good agreement with neutron-capture measurements \cite{Ruyl84}. Both the measurement of $\gamma$-ray energies 2618.0(2), 1497.1(2)\,keV and intensities are consistent with dominant $\beta$-feeding to the 4171.968(17)\,keV $3^+$ state.

The 1208\,keV transition was originally reported feeding a state in $^{50}$Ti with excitation energy 4171.8\,keV and a spin-parity of $3^+$ or $4^+$ (Ref. \cite{Alburger1984} and references therein). The same transition is reported here in addition to 5 weak feeding transitions observed at 1015, 1269, 1635, 1983 and 2065\,keV, decaying from states with {\it E$_x$} $\geq$ 5186\,keV. These are placed in the $^{50}$Ti level scheme for the first time (Figure \ref{fig_lvl_scheme}), feeding the 4171.9\,keV $3^+$ state and are observed in coincidence with the 1497 and 2618\,keV transitions. The newly-placed $\gamma$-ray transitions act to increase the $\gamma$ intensity populating the $3^+$ state by around 12$\%$.

The resulting direct $\beta$-feeding of the 4171.9\,keV $3^+$ level is found to be $\sim$0.06\%.
This result precludes direct feeding from the $^{50}$Sc $5^+$ ground state, where the corresponding log($ft$) ($\approx$~8) does not agree with the value expected for a $J$=2 transition ($\approx$~12) \cite{Singh1998}. 
It is unlikely that there is an overestimation of the $\beta$ branch intensity due to contributions from unobserved feeding. 
In such a scenario around 150 additional feeding transitions (assuming an upper limit for the detection of very weak transitions in the current work of $I_\gamma \leq$ 0.0004\% such as for the 1983\,keV $\gamma$ ray in Table \ref{table:intensities}) would be required to bring the experimental log($ft$) value into line with the typical range Ref. \cite{Singh1998}.
It is therefore much more likely that the 4171.9\,keV $3^+$ level is fed by the 257\,keV $2^+$ isomeric state in $^{50}$Sc ($T_{1/2}$ = 0.35(4) s).

A series of time-gated $\gamma$-ray addback energy spectra were produced using the decay portion of the long tape cycle. Each spectrum contained data integrated over 5~s with cuts placed from $\approx$ 0-300~s after the cessation of $^{50}$Ca implantation upon the tape. The intensity of the 2618\,keV peak was obtained as a function of time and compared to that of the neighboring 2675\,keV $\gamma$-ray energy sum peak (incorporating both the 1554 and 1121\,keV transitions). The decay curve gated on the 2675\,keV transition is consistent with the decay of the $^{50}$Sc ground state. The decay curve gated on the 2618\,keV transition, however, shows a significant decrease in intensity after around 100~s: around a factor of 2 lower compared to that of the 2675\,keV transition. Note that the contaminant 3132\,keV single-escape peak at 2618\,keV was subtracted from the data.

In summary, evidence was obtained in the current work for the direct feeding of the 4171.9\,keV $3^+$ level by the 257\,keV isomeric state in $^{50}$Sc.

\subsection{\label{ssec:lev_4310}The 4310\,keV state}

The excited state at 4310\,keV was known previously, for example it is reported as a 4322(20)\,keV state in the $^{49}$Ti(\textit{d,p}) stripping reaction \cite{Barnes65}. It was assigned a spin of \textit{J}=2 based on the \textit{L}=1 transfer strength observed following analysis of the differential cross section. The observation of $\gamma$ rays from this state feeding the 0$^{+}$ and $2_{1}^{+}$ states in a $^{49}$Ti(\textit{n,$\gamma$}) study \cite{Chen2019} also dissuades assignment of a large spin for this state.

The 2756 and 4308\,keV $\gamma$ rays are observed here in $\beta$ decay below a relative intensity of 0.01$\%$ with respect to the 2$_{1}^{+}$~$\rightarrow$~0$_{1}^{+}$ transition. The observed branching ratio of the weaker 4308\,keV transition is 26.8(35), slightly above the value reported in Ref. \cite{Chen2019} of 19.6(21). In the present study we have constrained the $\beta$ branching ratio to this state from the observation of a feeding transition with energy 1130\,keV. A $\gamma$-ray coincidence addback energy spectrum gated on this feeding transition is shown in Figure \ref{fig_energy2}.

An estimate of the branching ratio for the unobserved $2_2^+$\,(4310\,keV)~$\rightarrow$~$0_2^+$\,(3868\,keV) 440\,keV $E2$ transition is reported here via consideration of the 2314\,keV transition (also not observed in the current work) which is placed de-exciting the $0_2^+$ state \cite{Ruyl84}. We assume that the intensity of the 2314\,keV transition is directly analogous to the 440 since (i) no additional transitions are placed feeding the $0_{2}^{+}$ state, (ii) the $0_{2}^{+} \rightarrow 2_{1}^{+}$ decay has a branching ratio equal to unity and (iii) direct $\beta$ feeding to the $0_{2}^{+}$ state is unlikely. Analysis of the unobserved 2314\,keV transition focused on the spectrum shown in Figure \ref{fig_energy2} (a) gated on the 1130\,keV transition and centered on energies near 2314 and 2756\,keV. The background level observed near 2314\,keV yields an upper limit of $\approx$ 25 counts while around 450 counts are recorded in the same spectrum for the 2756\,keV peak. This corresponds to an intensity of $\leq$5$\%$ for the 2314\,keV transition relative to the 2756\,keV peak or equivalently for the 440\,keV transition, a $\leq$4$\%$ $\gamma$-ray branch from the 4310\,keV state. The likelihood of transitions with appreciable intensity populating the $0_2^+$ state from higher-lying levels seems low since the majority of these levels have been assigned spins with $J$ = 3 units of angular momentum and higher, with the exception of the 4310\,keV state. We also consider that a pure E0 transition from the $0_2^+$ state towards the ground state might compete with the 2314\,keV transition. However we estimate the magnitude of such a branch to be very small indeed, around 10$^{-6}$ of the intensity of the unobserved 2314\,keV E2 transition. The upper limit on the 440\,keV $\gamma$-ray branch from the 4310\,keV state could potentially be improved in future experiments via the use of bremsstrahlung- and Compton-suppressed clovers in GRIFFIN in addition to a higher time-integrated $^{50}$Ca beam intensity.

The difference in the observed intensity between the feeding and depopulating transitions yields a $\beta$ branch for this state of 0.002$\%$. This corresponds to a log({\it ft}) value around 12 for the {\it 5$^+$~$\rightarrow$~2$^+$} $\beta$ transition from the $^{50}$Sc ground state: well below the expected hindrance of this unique second-forbidden {\it $\Delta$J=3} transition \cite{Singh1998}. It is likely therefore that the obtained $\beta$ branch incorporates the {\it 2$^+$~$\rightarrow$~2$^+$} $\beta$ transition from the 257\,keV $^{50}$Sc isomer (yielding a log({\it ft}) of around 7). For this reason an upper limit is adopted for the $\beta$ branch given in Table \ref{table:BetaBR_LogFT}. Note that this value includes the intensity limit established for the unobserved 440\,keV $\gamma$ ray.

\subsection{The 5186\,keV state}
            
The 5186\,keV state was identified via the observation of $\gamma$ rays of 1015\,keV, 2511\,keV and 3631\,keV, which represent transitions to the $3^+$ member of the 4172\,keV doublet, $4^+_1$ 2675\,keV state and $2^+_1$ 1554\,keV state respectively. The relative branching ratios were determined using coincidence gates applied to the 1121 and 1554\,keV transitions and are in good agreement with previous $^{49}$Ti(\textit{n,$\gamma$}) measurements \cite{Chen2019}. This state was previously assigned a tentative spin of $J$=$(3)$ or $(4)$ based on the results of $\gamma$-ray angular distributions following the capture of polarized thermal neutrons on a polarized $^{49}$Ti target \cite{Ruyl84}. A positive parity was inferred from $L$=1 transfer observed with $^{49}$Ti(\textit{d,p}) \cite{Barnes65}. In the absence of any observed feeding transitions, a log({\it ft}) of 6.91(2) was obtained for this state from a $\beta$-branching ratio of $\approx$~0.1~$\%$, confirming the $4^+$ spin-parity assignment adopted in Table \ref{table:BetaBR_LogFT}.

\subsection{\label{ssec:lev_5548}The 5548\,keV state}

The 5548\,keV state was observed to decay by a single $\gamma$ ray of 3993\,keV with an intensity of $<$0.001$\%$. It appears in coincidence with the 1554\,keV transition to the $^{50}$Ti ground state. A $\gamma$-ray coincidence spectrum gated on the 1554\,keV transition is shown in Figure \ref{fig_energy2}, panel (c). The same spectrum without any coincidence requirements is shown in an inset to panel (c). While no $\gamma$ rays are observed to feed this state, two additional de-excitation branches were reported in Ref.\cite{Ruyl84} with energies of 2348.4 and 2872.8\,keV. The 2348.4\,keV $\gamma$ ray was tentatively placed feeding the 3199\,keV 6$^+$ state \cite{Ruyl84} and would therefore be expected to exhibit coincidences with the strongest transitions at 524, 1121 and 1554\,keV. No such coincidences have been observed in the current work. A transition with an energy of 2348.5(7)\,keV was also reported by Sona et al. \cite{Sona84} using $^{49}$Ti(d,p$\gamma$)$^{50}$Ti reactions but was placed feeding a 4172.3\,keV state, interpreted here as the (2)$^+$ 4172.5\,keV state belonging to the 4172\,keV doublet \cite{Chen2019}. As noted in section \ref{ssec:lev_4172}, only the 3$^+$ 4171.9\,keV state appears to be populated in the current work. The 2872.8\,keV $\gamma$ ray was reported with an appreciable intensity relative to the 3993\,keV branch. A possible explanation for the non-observation of this transition in the current work is that it is obscured by the Compton edges of more intense transitions, namely the 3132\,keV transition from the 5807\,keV state. The background level near the 3993\,keV transition is an order of magnitude lower despite a similar contribution from the Compton edge of the 4252\,keV transition. In addition, a coincidence gate placed on the 1554\,keV $\gamma$ ray does not exclude any of the most intense transitions or their associated backgrounds from incomplete energy collection.

The unobserved 2872.8\,keV $\gamma$ ray affects the determination of the $\beta$ branching ratio to the 5548\,keV state. Thus we have corrected the total decay intensity using the literature branching ratio in combination with the observed intensity in this study. This correction yields a $\beta$ branching ratio which is around a factor of two higher than the value obtained when no correction is applied. The calculated log($ft$) value decreases from $\approx$\,8.7 to 8.3 but is still at the upper limit of the tail of the $\Delta J$=$\pm1$ range of typical values found in Ref.\cite{Singh1998} 
Since the tentative 2348\,keV transition to the $6^+$ state at 3199\,keV is not confirmed in the present work it is possible that the 5548\,keV level has a spin of ($2,3$)$^+$ and is populated by the $^{50m}$Sc decay. Considering only the 3993 and the (unobserved) 2872.8\,keV transition intensities, a $\beta$-feeding intensity of 0.00137(16) is obtained corresponding to a log($ft$)=6.17(8) (See Table \ref{table:BetaBR_LogFT}).

\subsection{The 6123\,keV state}

Branching ratios of $\gamma$ rays de-exciting the 6123\,keV state with 1242, 1976, 2924 and 3448\,keV feeding the $5^+$ (4881\,keV), $4^+$ (4147\,keV), $6^+$ (3199\,keV) and $4^+$ (2675\,keV) states were established from $\gamma - \gamma$ coincidence gates placed at 1121\,keV to isolate the 2924 and 3448\,keV transitions, 1472\,keV for the 1976\,keV transition, and at 2206\,keV in order to obtain a value for the 1242\,keV transition. 

The data presented in Table \ref{table:intensities} are in reasonable agreement with results from neutron-capture measurements, where the 6123\,keV state was assigned {\it J$^{\pi}$=4$^+$} \cite{Ruyl84}, although the branching ratio of the 1976\,keV $\gamma$ ray is higher than the reported lower limit of $<15\%$ of the 1242\,keV transition. We note that the uncertainties provided in the literature are in general quite large for the weaker branches from the 6123\,keV state.

No evidence of the reported 1636\,keV transition is observed here, despite a significant intensity of 85~$\%$ relative to the 1242\,keV $\gamma$ ray reported in Ref. \cite{Ruyl84}. It is possible that the 1636\,keV $\gamma$ ray is obscured by the Compton scattered $6_1^+$~$\rightarrow$~$4_1^+$~$\rightarrow$~$2_1^+$ sum peak observed here at 1644\,keV. However, the 1242\,keV $\gamma$ ray is observed despite a 10-fold increase in background (dominated by backscattered 1554\,keV $\gamma$ rays) and only a modest increase in efficiency ($\approx$13$\%$). A search for additional gating transitions that might be used to isolate the unobserved 1636\,keV $\gamma$ ray (for example, the 2933 and 4487\,keV $\gamma$ rays which de-excite the 4487\,keV $2^+$ state) was not successful. We also note that the 1636\,keV $\gamma$ ray is not listed as unambiguously assigned in Table 3 of Ref.\cite{Ruyl84} and therefore no further corrections to the total decay intensity from the 6123\,keV state have been applied. No $\gamma$-ray transitions feeding this state were identified. The assigned spin and parity of $J^+$=$(4^+)$ is fully consistent with the extracted log({\it ft}) value in Table \ref{table:BetaBR_LogFT} of 5.56(4). However since it was not possible to confirm the 1636\,keV transition to the 4987\,keV, $2^+$ level in the present work then this spin and parity assignment remains tentative.

\subsection{The 6156\,keV state}

Four $\gamma$ rays of 1983, 2009\,keV, 3481\,keV and 4601\,keV are found to decay from the 6156\,keV state and populate the $3^+$ (4172\,keV), $4_2^+$ (4147\,keV), $4_1^+$ (2675\,keV) and $2_1^+$ (1554\,keV) states, respectively. No feeding $\gamma$-ray transitions were observed for this state. The 4601\,keV $\gamma$ ray was previously the only transition identified as de-exciting this state \cite{Ruyl84}, although here it is in fact the weakest branch at around 13$\%$ of the intensity of the 2009\,keV transition. The 3 remaining transitions are observed here following $\beta$ decay for the first time. A tentative range of spins {\it J$^{\pi}$=$(2,3,4)$} and positive parity were originally assigned to the 6156.0\,keV state in Ref.\cite{Ruyl84}. With the total decay intensity of the 4 de-exciting transitions taken into account, a $\beta$-branching ratio of $\approx$ 0.016$\%$ is obtained yielding a log({\it ft}) of 6.25(4), in agreement for {\it $\Delta$J=0,$\pm$1} transitions. A {\it J$^{\pi}$=4$^+$} assignment is therefore confirmed. It is noted that if the 4601\,keV $\gamma$ ray was indeed the only transition observed in the present experiment, then the transition rate would decrease by more than an order of magnitude (log({\it ft}) $\approx$ 7.4) but would remain in the tail of the typical {\it $\Delta$J=0,$\pm$1} values of Ref. \cite{Singh1998}.

\subsection{The 6237\,keV state}

The 6237.0\,keV state was identified from the observation of $\gamma$ rays of 2065\,keV feeding the $3^+$ member of the 4172\,keV doublet, 3038.0\,keV feeding the $6_1^+$ (3199\,keV) state and 3561.0\,keV populating the $4_1^+$ (2675.0\,keV) state. It is one of only two states observed in the present study that has not been reported elsewhere. The 3561\,keV $\gamma$ ray is observed in coincidence with the 1121\,keV and 1554\,keV transitions to the $2_1^+$ (1554\,keV) state and the ground state, respectively, whereas the 3038\,keV $\gamma$ ray is observed in coincidence with both of these transitions plus the 524\,keV $\gamma$ decay from the $6_1^+$ (3199\,keV) state. No $\gamma$-ray transitions were observed to feed this state. A log({\it ft}) value of 5.71(4) is calculated for this state as shown in Table \ref{table:BetaBR_LogFT}. A constraint is placed on the spin and parity of this state of $J^\pi$=$(4-5)^+$ based on the log({\it ft}) value and the observation of the de-populating 2065\,keV transition feeding the $3^+$ member of the 4172\,keV doublet.

\subsection{The 6625\,keV state}	

The 6625.0\,keV state has not been observed in any previous work and was identified here via the observation of $\gamma$ rays at 3427\,keV decaying to the $6_1^+$ (3199\,keV) state and 3950\,keV to the $4_1^+$ (2675\,keV) state. The 3950\,keV $\gamma$ ray is observed in coincidence with the 1121\,keV and 1554\,keV transitions to the $2_1^+$ and $0_1^+$ states, respectively, while for the 3427\,keV $\gamma$ ray an additional strong coincidence is found with the 524\,keV transition to the $4_1^+$ state. No feeding transitions were observed, although the excitation energy lies close to the $\beta$-decay $Q$ value (similarly for the 6237\,keV state). A log({\it ft}) value of 5.35(9) is calculated and from this an assignment of a positive parity with a spin value in the range $J$=$(4-6)$ is provided in Table \ref{table:BetaBR_LogFT}.

Several excited states in $^{50}$Ti with $E_x$~$\approx$\,6.3-6.8\,MeV are proposed in Ref. \cite{Ruyl84} with (tentatively) assigned spin and parities which would mean they are accessible via allowed Gamow-Teller $\beta$ transitions. A search was performed of the data for the associated decay $\gamma$ rays including examining $\gamma$-$\gamma$-$\gamma$ triple coincidences between the 524, 1121 or 1554\,keV transitions to reduce the background contribution from Compton scattered sum peaks which may obscure very weak transitions. No additional $\gamma$ rays were identified. The 6625.0\,keV state represents the highest observed excitation energy populated in the $\beta ^-$ decay of the $^{50}$Sc.

\subsection{Shell model calculations}

The stable $^{50}$Ti nucleus sits just above doubly-magic $^{48}$Ca with a closed-shell of 28 neutrons and 2 valence protons outside the magic proton shell closure of 20. It is expected that the spherical shell model will reproduce the excitations in $^{50}$Ti with good accuracy. Shell model calculations using both phenomenological and {\it ab initio} based interactions have been performed for comparison with the experimental data.

Shell model calculations were performed with the NuShellX@MSU shell-model code \cite{Brown2014} using the phenomenological KB3G \cite{Poves01} and GXPF1A \cite{Honma2005} interactions in the $pf$ valence space ($0f_{7/2}$, $1p_{3/2}$, $0f_{5/2}$, $1p_{1/2}$), known to well reproduce experimental data in this region.  In addition, we derive {\it ab initio} shell-model Hamiltonians within the valence-space in-medium similarity renormalization group (VS-IMSRG) framework \cite{Tsuk12SM,Stroberg:2019mxo, Bogn14SM,Stro17ENO,Stro16TNO}, based on two-nucleon (NN) and three-nucleon (3N) forces derived from chiral effective field theory \cite{Epel09RMP,Mach11PR}. 
The details of the particular input NN+3N interaction used here (EM1.8\textunderscore2.0), and developed in Refs.~\cite{Hebe11fits,Simo16unc,Simo17SatFinNuc}, as well as the specifics of the Hamiltonian are described in Ref. \cite{Garnsworthy_2017B}.

Starting in a single-particle spherical harmonic oscillator (HO) basis with energy $\hbar\omega=16$\,MeV, we first transform the input Hamiltonian to the Hartree-Fock (HF) basis, then use the Magnus formulation of the VS-IMSRG \cite{Morr15Magnus,Herg16PR}, with the ensemble normal ordering procedure~\cite{Stro17ENO}, which captures the bulk effects of residual 3N forces among valence nucleons, to produce an approximate unitary transformation which decouples the $^{40}$Ca core.  A second transformation is performed to decouple a specific $pf$-shell valence-space Hamiltonian appropriate for $^{50}$Ti. These results are well converged within the basis size $e=2n+l \le e_{\mathrm{max}}=12$ and $e_1 + e_2 + e_3 \le E_{\mathrm{3max}} = 16 $. We finally use the approximate unitary transformation to decouple an effective valence-space $E2$ or $M1$ operator consistent with the valence-space Hamiltonian \cite{Parz:prc2017}.

\begin{table}[t]
\begin{centering}
\caption{\label{Tab:calc}Transition strength, $B(E2)$, values are given in $e^2fm^4$ and quadrupole moments, $Q$, are given in $e^2fm^2$. Experimental data are taken from Ref. \cite{Chen2019}.}
\begin{tabular}{cc|ccc}
\hline
 & Exp. & KB3G & GXPF1A & VS-IM-SRG \\
\hline
$B(E2: 2_1-0_1)$ &	58(2)	&	99.9	&	100.4	&	38.2	\\
$B(E2: 4_1-2_1)$ &	60(13)	&	98.2	&	98.0	&	36.1	\\
$B(E2: 6_1-4_1)$ &	34(1)	&	46.7	&	47.0	&	16.9	\\
$B(E2: 2_{2p2h}-0_{2p2h})$ &	-	&	10.1	&	39.5	&	26.7	\\
\hline
$Q (2_1)$ & 8(16) & 6.77 & 6.53 & 4.69 \\
$Q (2_2)$ & - & -19.61 & -12.77 & -11.36 \\
$Q (2_3)$ & - & 13.51 & 13.30 & 9.76 \\
\hline
\end{tabular}
\end{centering}
\end{table}%

Three distinct sets of states emerge from the calculations. Firstly the ground-state band where a pair of protons occupy primarily the $\pi0f_{7/2}$ orbital ($J^\pi = 0^+, 2^+, 4^+, 6^+$) and the neutron orbitals are filled up to the $N=28$ shell gap, secondly a neutron 1-particle-1-hole ($\nu 0f_{7/2}^{-1}-1p_{3/2}^{1}$) configuration ($J^\pi = 2^+, 3^+, 4^+, 5^+$) and finally a neutron 2-particle-2-hole ($\nu 0f_{7/2}^{-2}-1p_{3/2}^{2}$) configuration ($J^\pi = 0^+, 2^+$).
In both of these cross-shell excitations the particles occupy primarily the $1p_{3/2}$ orbital above $N=28$. A detailed comparison of every experimentally observed excited state is difficult due to the level density and in some cases the lack of firm spin assignments. Instead we concentrate here on an examination of these three low-lying structures.

\begin{figure*}[ht]
		\includegraphics[angle=0,width=0.95\linewidth]{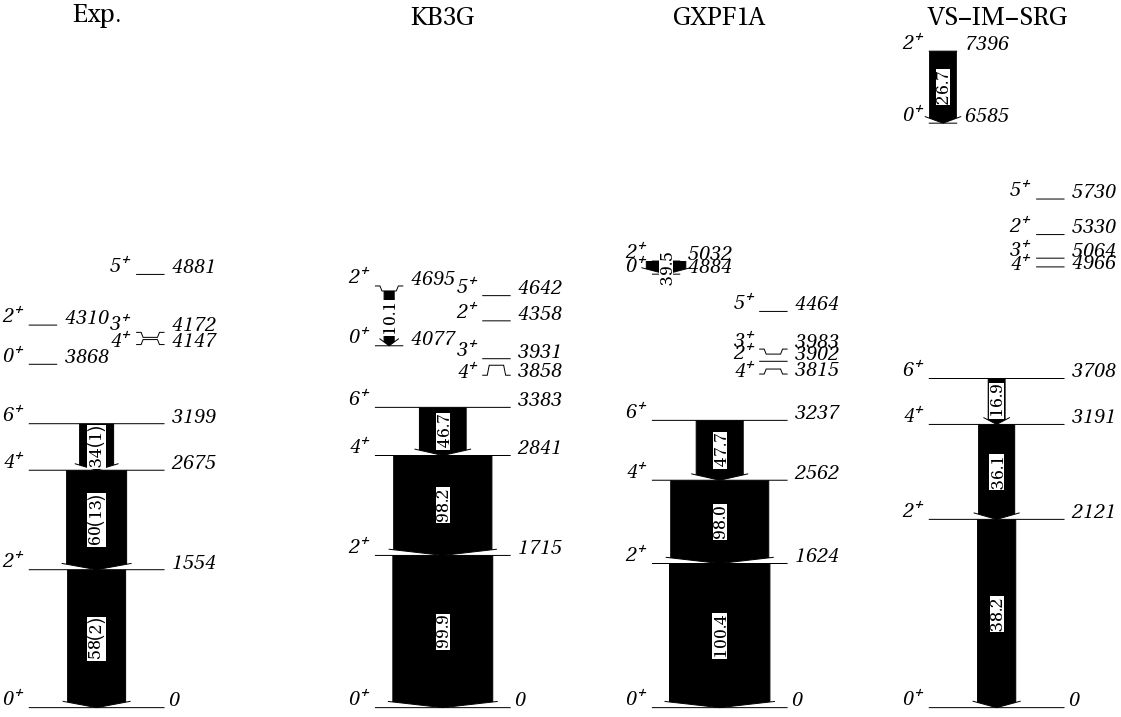}
		\centering
        \caption{Selected experimentally observed levels in $^{50}$Ti are compared to the results of three different theory calculations. Excited states belonging to the ground state band (full-width levels), $2p2h$ (quarter-width, left) and $1p1h$ particle-hole excitations (quarter-width, right) are included. {\it B(E2)} transition strengths are provided in units of $e^2fm^4$ and are proportional to the widths of the arrows.}
        \label{fig_transition_strength}
		\end{figure*}

Select properties from the three calculations are shown in Figure \ref{fig_transition_strength} and are given in Table \ref{Tab:calc} in comparison to the experimental data. While the first excited {\it 2$^+$} energy predicted in the VS-IM-SRG is several hundred keV higher than experiment, possibly arising from a too-large shell gap, the spacing between higher-lying states agrees well with experiment.

The degree of collectivity as evidenced in the magnitude of the $B(E2)$ values is also underpredicted by the VS-IM-SRG. This is expected because of the limitations placed on the model space used for the transformation. Despite this, the spectroscopic information such as the pattern of $B(E2)$ values and quadrupole moments ($Q$) are well reproduced, as was first noted in Ref. \cite{Henderson:2017dqc}. 

All three calculations predict a positive quadrupole moment in the ground state and $2p-2h$ bands while the $1p-1h$ configuration is predicted to be negative, all with similar magnitudes. It is interesting that the $B(E2)$ values predicted for the ground-state band are consistent between the interactions but that the predictions for the $2p-2h$ band are quite different, ranging from 10 to 70\% that of the $2^+_1\rightarrow 0^+_1$ transition. This distinction might be used in the future to distinguish between the calculations once an experimental value for the strength of this transition becomes available.

\section{\label{sec:conclusion}Conclusion}


The GRIFFIN $\gamma$-ray spectrometer at the TRIUMF-ISAC facility has been used to study the $\beta$ decay of $^{50}$Sc to $^{50}$Ti using a radioactive beam of $^{50}$Ca. $\beta$-decay branching ratios from both the ground state and isomeric state of $^{50}$Sc to 16 excited states in $^{50}$Ti are determined from a total of 38 $\gamma$-ray transitions which significantly expands the information available over previous studies. 
Relative intensities are measured to less than 0.001{\it $\%$} that of the strongest transition with the majority of $\gamma$-ray transitions observed here in $\beta$ decay for the first time. 
The data are compared to shell-model calculations utilizing both phenomenologically-derived interactions employed in the {\it pf} shell as well as a state-of-the-art, {\it ab initio} based interaction built in the valence-space in-medium similarity renormalization group framework. The differences in predictions for cross-shell excitations across the $N=28$ neutron shell closure are discussed.

\begin{acknowledgments}
We would like to thank the operations and beam delivery staff at TRIUMF for providing the radioactive beam. C.E.S. acknowledges support from the Canada Research Chairs program. The GRIFFIN spectrometer was jointly funded by the Canadian Foundation for Innovation (CFI), TRIUMF, and the University of Guelph. TRIUMF receives federal funding via a contribution agreement through the National Research Council Canada (NRC). This work was supported in part by the Natural Sciences and Engineering Research Council of Canada (NSERC). Figure 3 of this article has been created using the SciDraw scientific figure preparation system \cite{Caprio2005}.
\end{acknowledgments}

\bibliographystyle{apsrev}
\bibliography{main}

\providecommand{\BIBUlf}{Ulf-G}
\begin{thebibliography}{48}
\expandafter\ifx\csname natexlab\endcsname\relax\def\natexlab#1{#1}\fi
\expandafter\ifx\csname bibnamefont\endcsname\relax
  \def\bibnamefont#1{#1}\fi
\expandafter\ifx\csname bibfnamefont\endcsname\relax
  \def\bibfnamefont#1{#1}\fi
\expandafter\ifx\csname citenamefont\endcsname\relax
  \def\citenamefont#1{#1}\fi
\expandafter\ifx\csname url\endcsname\relax
  \def\url#1{\texttt{#1}}\fi
\expandafter\ifx\csname urlprefix\endcsname\relax\def\urlprefix{URL }\fi
\providecommand{\bibinfo}[2]{#2}
\providecommand{\eprint}[2][]{\url{#2}}

\bibitem[{\citenamefont{Heyde et~al.}(1982)\citenamefont{Heyde, Van~Isacker,
  Waroquier, Wenes, and Sambataro}}]{Heyde1982}
\bibinfo{author}{\bibfnamefont{K.}~\bibnamefont{Heyde}},
  \bibinfo{author}{\bibfnamefont{P.}~\bibnamefont{Van~Isacker}},
  \bibinfo{author}{\bibfnamefont{M.}~\bibnamefont{Waroquier}},
  \bibinfo{author}{\bibfnamefont{G.}~\bibnamefont{Wenes}}, \bibnamefont{and}
  \bibinfo{author}{\bibfnamefont{M.}~\bibnamefont{Sambataro}},
  \bibinfo{journal}{Phys. Rev. C} \textbf{\bibinfo{volume}{25}},
  \bibinfo{pages}{3160} (\bibinfo{year}{1982}),
  \urlprefix\url{https://link.aps.org/doi/10.1103/PhysRevC.25.3160}.

\bibitem[{\citenamefont{Heyde and Wood}(2011)}]{Heyde2011}
\bibinfo{author}{\bibfnamefont{K.}~\bibnamefont{Heyde}} \bibnamefont{and}
  \bibinfo{author}{\bibfnamefont{J.~L.} \bibnamefont{Wood}},
  \bibinfo{journal}{Reviews of Modern Physics} \textbf{\bibinfo{volume}{83}},
  \bibinfo{pages}{1467} (\bibinfo{year}{2011}),
  \urlprefix\url{http://link.aps.org/doi/10.1103/RevModPhys.83.1467}.

\bibitem[{\citenamefont{Epelbaum et~al.}(2009)\citenamefont{Epelbaum, Hammer,
  and Mei{\ss}ner}}]{Epel09RMP}
\bibinfo{author}{\bibfnamefont{E.}~\bibnamefont{Epelbaum}},
  \bibinfo{author}{\bibfnamefont{H.~W.} \bibnamefont{Hammer}},
  \bibnamefont{and}
  \bibinfo{author}{\bibfnamefont{{\BIBUlf}.}~\bibnamefont{Mei{\ss}ner}},
  \bibinfo{journal}{Rev. Mod. Phys.} \textbf{\bibinfo{volume}{81}},
  \bibinfo{pages}{1773} (\bibinfo{year}{2009}).

\bibitem[{\citenamefont{Machleidt and Entem}(2011)}]{Mach11PR}
\bibinfo{author}{\bibfnamefont{R.}~\bibnamefont{Machleidt}} \bibnamefont{and}
  \bibinfo{author}{\bibfnamefont{D.~R.} \bibnamefont{Entem}},
  \bibinfo{journal}{Phys. Rep.} \textbf{\bibinfo{volume}{503}},
  \bibinfo{pages}{1} (\bibinfo{year}{2011}).

\bibitem[{\citenamefont{Bogner et~al.}(2007)\citenamefont{Bogner, Furnstahl,
  and Perry}}]{Bogn07SRG}
\bibinfo{author}{\bibfnamefont{S.~K.} \bibnamefont{Bogner}},
  \bibinfo{author}{\bibfnamefont{R.~J.} \bibnamefont{Furnstahl}},
  \bibnamefont{and} \bibinfo{author}{\bibfnamefont{R.~J.} \bibnamefont{Perry}},
  \bibinfo{journal}{Phys. Rev. C} \textbf{\bibinfo{volume}{75}},
  \bibinfo{pages}{061001(R)} (\bibinfo{year}{2007}).

\bibitem[{\citenamefont{Bogner et~al.}(2010)\citenamefont{Bogner, Furnstahl,
  and Schwenk}}]{Bogn10PPNP}
\bibinfo{author}{\bibfnamefont{S.~K.} \bibnamefont{Bogner}},
  \bibinfo{author}{\bibfnamefont{R.~J.} \bibnamefont{Furnstahl}},
  \bibnamefont{and} \bibinfo{author}{\bibfnamefont{A.}~\bibnamefont{Schwenk}},
  \bibinfo{journal}{Prog. Part. Nucl. Phys.} \textbf{\bibinfo{volume}{65}},
  \bibinfo{pages}{94} (\bibinfo{year}{2010}).

\bibitem[{\citenamefont{Hebeler et~al.}(2015)\citenamefont{Hebeler, Holt,
  Men{\'e}ndez, and Schwenk}}]{Hebe15ARNPS}
\bibinfo{author}{\bibfnamefont{K.}~\bibnamefont{Hebeler}},
  \bibinfo{author}{\bibfnamefont{J.~D.} \bibnamefont{Holt}},
  \bibinfo{author}{\bibfnamefont{J.}~\bibnamefont{Men{\'e}ndez}},
  \bibnamefont{and} \bibinfo{author}{\bibfnamefont{A.}~\bibnamefont{Schwenk}},
  \bibinfo{journal}{Ann. Rev. Nucl. Part. Sci.} \textbf{\bibinfo{volume}{65}},
  \bibinfo{pages}{457} (\bibinfo{year}{2015}).

\bibitem[{\citenamefont{Rowe and Wood}(2018)}]{rowanwood}
\bibinfo{author}{\bibfnamefont{D.~J.} \bibnamefont{Rowe}} \bibnamefont{and}
  \bibinfo{author}{\bibfnamefont{J.~L.} \bibnamefont{Wood}},
  \emph{\bibinfo{title}{Fundamentals Of Nuclear Models - Volume 2: Unified
  Models}} (\bibinfo{publisher}{World Scientific Publishing Co Pte Ltd},
  \bibinfo{address}{Singapore}, \bibinfo{year}{2018}).

\bibitem[{\citenamefont{Sorlin and Porquet}(2013)}]{Sorlin12}
\bibinfo{author}{\bibfnamefont{O.}~\bibnamefont{Sorlin}} \bibnamefont{and}
  \bibinfo{author}{\bibfnamefont{M.-G.} \bibnamefont{Porquet}},
  \bibinfo{journal}{Phys. Scr.} \textbf{\bibinfo{volume}{T152}},
  \bibinfo{pages}{014003} (\bibinfo{year}{2013}).

\bibitem[{\citenamefont{Poves et~al.}(2001)\citenamefont{Poves,
  S\'{a}nchez-Solano, Caurier, and Nowacki}}]{Poves01}
\bibinfo{author}{\bibfnamefont{A.}~\bibnamefont{Poves}},
  \bibinfo{author}{\bibfnamefont{J.}~\bibnamefont{S\'{a}nchez-Solano}},
  \bibinfo{author}{\bibfnamefont{E.}~\bibnamefont{Caurier}}, \bibnamefont{and}
  \bibinfo{author}{\bibfnamefont{F.}~\bibnamefont{Nowacki}},
  \bibinfo{journal}{Nuclear Physics A} \textbf{\bibinfo{volume}{694}},
  \bibinfo{pages}{157 } (\bibinfo{year}{2001}), ISSN \bibinfo{issn}{0375-9474},
  \urlprefix\url{http://www.sciencedirect.com/science/article/pii/S0375947401009678}.

\bibitem[{\citenamefont{Svensson and Garnsworthy}(2013)}]{Svensson2013}
\bibinfo{author}{\bibfnamefont{C.~E.} \bibnamefont{Svensson}} \bibnamefont{and}
  \bibinfo{author}{\bibfnamefont{A.~B.} \bibnamefont{Garnsworthy}},
  \bibinfo{journal}{Hyperfine Interact.} \textbf{\bibinfo{volume}{225}},
  \bibinfo{pages}{127} (\bibinfo{year}{2013}), ISSN \bibinfo{issn}{0304-3843}.

\bibitem[{\citenamefont{Rizwan et~al.}(2016)\citenamefont{Rizwan, Garnsworthy,
  Andreoiu, Ball, Chester, Domingo, Dunlop, Hackman, Rand, Smith
  et~al.}}]{Rizwan2016}
\bibinfo{author}{\bibfnamefont{U.}~\bibnamefont{Rizwan}},
  \bibinfo{author}{\bibfnamefont{A.~B.} \bibnamefont{Garnsworthy}},
  \bibinfo{author}{\bibfnamefont{C.}~\bibnamefont{Andreoiu}},
  \bibinfo{author}{\bibfnamefont{G.~C.} \bibnamefont{Ball}},
  \bibinfo{author}{\bibfnamefont{A.}~\bibnamefont{Chester}},
  \bibinfo{author}{\bibfnamefont{T.}~\bibnamefont{Domingo}},
  \bibinfo{author}{\bibfnamefont{R.}~\bibnamefont{Dunlop}},
  \bibinfo{author}{\bibfnamefont{G.}~\bibnamefont{Hackman}},
  \bibinfo{author}{\bibfnamefont{E.~T.} \bibnamefont{Rand}},
  \bibinfo{author}{\bibfnamefont{J.~K.} \bibnamefont{Smith}},
  \bibnamefont{et~al.}, \bibinfo{journal}{Nucl. Instruments Methods Phys. Res.
  Sect. A Accel. Spectrometers, Detect. Assoc. Equip.}
  \textbf{\bibinfo{volume}{820}}, \bibinfo{pages}{126} (\bibinfo{year}{2016}),
  ISSN \bibinfo{issn}{01689002}.

\bibitem[{\citenamefont{Garnsworthy
  et~al.}(2017{\natexlab{a}})\citenamefont{Garnsworthy, Pearson, Bishop, Shaw,
  Smith, Bowry, Bildstein, Hackman, Garrett, Linn et~al.}}]{Garnsworthy2017A}
\bibinfo{author}{\bibfnamefont{A.~B.} \bibnamefont{Garnsworthy}},
  \bibinfo{author}{\bibfnamefont{C.~J.} \bibnamefont{Pearson}},
  \bibinfo{author}{\bibfnamefont{D.}~\bibnamefont{Bishop}},
  \bibinfo{author}{\bibfnamefont{B.}~\bibnamefont{Shaw}},
  \bibinfo{author}{\bibfnamefont{J.~K.} \bibnamefont{Smith}},
  \bibinfo{author}{\bibfnamefont{M.}~\bibnamefont{Bowry}},
  \bibinfo{author}{\bibfnamefont{V.}~\bibnamefont{Bildstein}},
  \bibinfo{author}{\bibfnamefont{G.}~\bibnamefont{Hackman}},
  \bibinfo{author}{\bibfnamefont{P.~E.} \bibnamefont{Garrett}},
  \bibinfo{author}{\bibfnamefont{Y.}~\bibnamefont{Linn}}, \bibnamefont{et~al.},
  \bibinfo{journal}{Nuclear Instruments and Methods in Physics Research Section
  A: Accelerators, Spectrometers, Detectors and Associated Equipment}
  \textbf{\bibinfo{volume}{853}}, \bibinfo{pages}{85 }
  (\bibinfo{year}{2017}{\natexlab{a}}), ISSN \bibinfo{issn}{0168-9002},
  \urlprefix\url{http://www.sciencedirect.com/science/article/pii/S0168900217302243}.

\bibitem[{\citenamefont{Garnsworthy et~al.}(2019)\citenamefont{Garnsworthy,
  Svensson, Bowry, Dunlop, MacLean, Olaizola, Smith, Ali, Andreoiu, Ash
  et~al.}}]{Garnsworthy2018}
\bibinfo{author}{\bibfnamefont{A.~B.} \bibnamefont{Garnsworthy}},
  \bibinfo{author}{\bibfnamefont{C.~E.} \bibnamefont{Svensson}},
  \bibinfo{author}{\bibfnamefont{M.}~\bibnamefont{Bowry}},
  \bibinfo{author}{\bibfnamefont{R.}~\bibnamefont{Dunlop}},
  \bibinfo{author}{\bibfnamefont{A.~D.} \bibnamefont{MacLean}},
  \bibinfo{author}{\bibfnamefont{B.}~\bibnamefont{Olaizola}},
  \bibinfo{author}{\bibfnamefont{J.~K.} \bibnamefont{Smith}},
  \bibinfo{author}{\bibfnamefont{F.~A.} \bibnamefont{Ali}},
  \bibinfo{author}{\bibfnamefont{C.}~\bibnamefont{Andreoiu}},
  \bibinfo{author}{\bibfnamefont{J.~E.} \bibnamefont{Ash}},
  \bibnamefont{et~al.}, \bibinfo{journal}{Nuclear Instruments and Methods in
  Physics Research Section A: Accelerators, Spectrometers, Detectors and
  Associated Equipment} \textbf{\bibinfo{volume}{918}}, \bibinfo{pages}{9 }
  (\bibinfo{year}{2019}), ISSN \bibinfo{issn}{0168-9002},
  \urlprefix\url{http://www.sciencedirect.com/science/article/pii/S0168900218317662}.

\bibitem[{\citenamefont{Garnsworthy
  et~al.}(2017{\natexlab{b}})\citenamefont{Garnsworthy, Bowry, Olaizola, Holt,
  Stroberg, Cruz, Georges, Hackman, MacLean, Measures
  et~al.}}]{Garnsworthy_2017B}
\bibinfo{author}{\bibfnamefont{A.~B.} \bibnamefont{Garnsworthy}},
  \bibinfo{author}{\bibfnamefont{M.}~\bibnamefont{Bowry}},
  \bibinfo{author}{\bibfnamefont{B.}~\bibnamefont{Olaizola}},
  \bibinfo{author}{\bibfnamefont{J.~D.} \bibnamefont{Holt}},
  \bibinfo{author}{\bibfnamefont{S.~R.} \bibnamefont{Stroberg}},
  \bibinfo{author}{\bibfnamefont{S.}~\bibnamefont{Cruz}},
  \bibinfo{author}{\bibfnamefont{S.}~\bibnamefont{Georges}},
  \bibinfo{author}{\bibfnamefont{G.}~\bibnamefont{Hackman}},
  \bibinfo{author}{\bibfnamefont{A.~D.} \bibnamefont{MacLean}},
  \bibinfo{author}{\bibfnamefont{J.}~\bibnamefont{Measures}},
  \bibnamefont{et~al.}, \bibinfo{journal}{Phys. Rev. C}
  \textbf{\bibinfo{volume}{96}}, \bibinfo{pages}{044329}
  (\bibinfo{year}{2017}{\natexlab{b}}),
  \urlprefix\url{https://link.aps.org/doi/10.1103/PhysRevC.96.044329}.

\bibitem[{\citenamefont{Warburton et~al.}(1970)\citenamefont{Warburton,
  Alburger, and Engelbertink}}]{Warburton1970}
\bibinfo{author}{\bibfnamefont{E.~K.} \bibnamefont{Warburton}},
  \bibinfo{author}{\bibfnamefont{D.~E.} \bibnamefont{Alburger}},
  \bibnamefont{and} \bibinfo{author}{\bibfnamefont{G.~A.~P.}
  \bibnamefont{Engelbertink}}, \bibinfo{journal}{Phys. Rev. C}
  \textbf{\bibinfo{volume}{2}}, \bibinfo{pages}{1427} (\bibinfo{year}{1970}).

\bibitem[{\citenamefont{Bylinskii and Craddock}(2013)}]{Bylinskii2013}
\bibinfo{author}{\bibfnamefont{I.}~\bibnamefont{Bylinskii}} \bibnamefont{and}
  \bibinfo{author}{\bibfnamefont{M.~K.} \bibnamefont{Craddock}},
  \bibinfo{journal}{Hyperfine Interact.} \textbf{\bibinfo{volume}{225}},
  \bibinfo{pages}{9} (\bibinfo{year}{2013}), ISSN \bibinfo{issn}{0304-3843}.

\bibitem[{\citenamefont{Langevin et~al.}(1983)\citenamefont{Langevin, D{\'
  e}traz, Guillemaud-Mueller, Mueller, Thibault, Touchard, Klotz, Mieh{\' e},
  Walter, Epherre et~al.}}]{Langevin1983}
\bibinfo{author}{\bibfnamefont{M.}~\bibnamefont{Langevin}},
  \bibinfo{author}{\bibfnamefont{C.}~\bibnamefont{D{\' e}traz}},
  \bibinfo{author}{\bibfnamefont{D.}~\bibnamefont{Guillemaud-Mueller}},
  \bibinfo{author}{\bibfnamefont{A.}~\bibnamefont{Mueller}},
  \bibinfo{author}{\bibfnamefont{C.}~\bibnamefont{Thibault}},
  \bibinfo{author}{\bibfnamefont{F.}~\bibnamefont{Touchard}},
  \bibinfo{author}{\bibfnamefont{G.}~\bibnamefont{Klotz}},
  \bibinfo{author}{\bibfnamefont{C.}~\bibnamefont{Mieh{\' e}}},
  \bibinfo{author}{\bibfnamefont{G.}~\bibnamefont{Walter}},
  \bibinfo{author}{\bibfnamefont{M.}~\bibnamefont{Epherre}},
  \bibnamefont{et~al.}, \bibinfo{journal}{Physics Letters B}
  \textbf{\bibinfo{volume}{130}}, \bibinfo{pages}{251 } (\bibinfo{year}{1983}),
  ISSN \bibinfo{issn}{0370-2693}.

\bibitem[{\citenamefont{Basu and Sonzogni}(2013)}]{Basu2013}
\bibinfo{author}{\bibfnamefont{S.~K.} \bibnamefont{Basu}} \bibnamefont{and}
  \bibinfo{author}{\bibfnamefont{A.~A.} \bibnamefont{Sonzogni}},
  \bibinfo{journal}{Nucl. Data Sheets} \textbf{\bibinfo{volume}{114}},
  \bibinfo{pages}{435} (\bibinfo{year}{2013}).

\bibitem[{\citenamefont{Haenni and Sugihara}(1977)}]{Haenni1977}
\bibinfo{author}{\bibfnamefont{D.~R.} \bibnamefont{Haenni}} \bibnamefont{and}
  \bibinfo{author}{\bibfnamefont{T.~T.} \bibnamefont{Sugihara}},
  \bibinfo{journal}{Phys. Rev. C} \textbf{\bibinfo{volume}{16}},
  \bibinfo{pages}{1129} (\bibinfo{year}{1977}),
  \urlprefix\url{https://link.aps.org/doi/10.1103/PhysRevC.16.1129}.

\bibitem[{\citenamefont{Alburger et~al.}(1984)\citenamefont{Alburger,
  Warburton, and Brown}}]{Alburger1984}
\bibinfo{author}{\bibfnamefont{D.~E.} \bibnamefont{Alburger}},
  \bibinfo{author}{\bibfnamefont{E.~K.} \bibnamefont{Warburton}},
  \bibnamefont{and} \bibinfo{author}{\bibfnamefont{B.~A.} \bibnamefont{Brown}},
  \bibinfo{journal}{Phys. Rev. C} \textbf{\bibinfo{volume}{30}},
  \bibinfo{pages}{1005} (\bibinfo{year}{1984}),
  \urlprefix\url{https://link.aps.org/doi/10.1103/PhysRevC.30.1005}.

\bibitem[{\citenamefont{Ruyl et~al.}(1984)\citenamefont{Ruyl, Haas, Endt, and
  Zybert}}]{Ruyl84}
\bibinfo{author}{\bibfnamefont{J.~F. A.~G.} \bibnamefont{Ruyl}},
  \bibinfo{author}{\bibfnamefont{J.~B. M.~D.} \bibnamefont{Haas}},
  \bibinfo{author}{\bibfnamefont{P.~M.} \bibnamefont{Endt}}, \bibnamefont{and}
  \bibinfo{author}{\bibfnamefont{L.}~\bibnamefont{Zybert}},
  \bibinfo{journal}{Nuclear Physics A} \textbf{\bibinfo{volume}{419}},
  \bibinfo{pages}{439 } (\bibinfo{year}{1984}), ISSN \bibinfo{issn}{0375-9474},
  \urlprefix\url{http://www.sciencedirect.com/science/article/pii/0375947484906262}.

\bibitem[{\citenamefont{Kulp et~al.}(2007)\citenamefont{Kulp, Wood, Allmond,
  Eimer, Furse, Krane, Loats, Schmelzenbach, Stapels, Larimer et~al.}}]{Kulp07}
\bibinfo{author}{\bibfnamefont{W.~D.} \bibnamefont{Kulp}},
  \bibinfo{author}{\bibfnamefont{J.~L.} \bibnamefont{Wood}},
  \bibinfo{author}{\bibfnamefont{J.~M.} \bibnamefont{Allmond}},
  \bibinfo{author}{\bibfnamefont{J.}~\bibnamefont{Eimer}},
  \bibinfo{author}{\bibfnamefont{D.}~\bibnamefont{Furse}},
  \bibinfo{author}{\bibfnamefont{K.~S.} \bibnamefont{Krane}},
  \bibinfo{author}{\bibfnamefont{J.}~\bibnamefont{Loats}},
  \bibinfo{author}{\bibfnamefont{P.}~\bibnamefont{Schmelzenbach}},
  \bibinfo{author}{\bibfnamefont{C.~J.} \bibnamefont{Stapels}},
  \bibinfo{author}{\bibfnamefont{R.-M.} \bibnamefont{Larimer}},
  \bibnamefont{et~al.}, \bibinfo{journal}{Phys. Rev. C}
  \textbf{\bibinfo{volume}{76}}, \bibinfo{pages}{034319}
  (\bibinfo{year}{2007}),
  \urlprefix\url{https://link.aps.org/doi/10.1103/PhysRevC.76.034319}.

\bibitem[{\citenamefont{Garrett et~al.}(2012)\citenamefont{Garrett, Bangay,
  Diaz~Varela, Ball, Cross, Demand, Finlay, Garnsworthy, Green, Hackman
  et~al.}}]{Garrett12}
\bibinfo{author}{\bibfnamefont{P.~E.} \bibnamefont{Garrett}},
  \bibinfo{author}{\bibfnamefont{J.}~\bibnamefont{Bangay}},
  \bibinfo{author}{\bibfnamefont{A.}~\bibnamefont{Diaz~Varela}},
  \bibinfo{author}{\bibfnamefont{G.~C.} \bibnamefont{Ball}},
  \bibinfo{author}{\bibfnamefont{D.~S.} \bibnamefont{Cross}},
  \bibinfo{author}{\bibfnamefont{G.~A.} \bibnamefont{Demand}},
  \bibinfo{author}{\bibfnamefont{P.}~\bibnamefont{Finlay}},
  \bibinfo{author}{\bibfnamefont{A.~B.} \bibnamefont{Garnsworthy}},
  \bibinfo{author}{\bibfnamefont{K.~L.} \bibnamefont{Green}},
  \bibinfo{author}{\bibfnamefont{G.}~\bibnamefont{Hackman}},
  \bibnamefont{et~al.}, \bibinfo{journal}{Phys. Rev. C}
  \textbf{\bibinfo{volume}{86}}, \bibinfo{pages}{044304}
  (\bibinfo{year}{2012}),
  \urlprefix\url{https://link.aps.org/doi/10.1103/PhysRevC.86.044304}.

\bibitem[{\citenamefont{Chen and Singh}(2019)}]{Chen2019}
\bibinfo{author}{\bibfnamefont{J.}~\bibnamefont{Chen}} \bibnamefont{and}
  \bibinfo{author}{\bibfnamefont{B.}~\bibnamefont{Singh}},
  \bibinfo{journal}{Nucl. Data Sheets} \textbf{\bibinfo{volume}{157}},
  \bibinfo{pages}{1} (\bibinfo{year}{2019}).

\bibitem[{\citenamefont{Kibe\'di et~al.}(2008)}]{Kibedi2008}
\bibinfo{author}{\bibfnamefont{T.}~\bibnamefont{Kibe\'di}}
  \bibnamefont{et~al.}, \bibinfo{journal}{Nucl. Instruments Methods Phys. Res.
  Sect. A Accel. Spectrometers, Detect. Assoc. Equip.}
  \textbf{\bibinfo{volume}{589}}, \bibinfo{pages}{202} (\bibinfo{year}{2008}).

\bibitem[{\citenamefont{Singh et~al.}(1998)\citenamefont{Singh, Rodriguez,
  Wong, and Tuli}}]{Singh1998}
\bibinfo{author}{\bibfnamefont{B.}~\bibnamefont{Singh}},
  \bibinfo{author}{\bibfnamefont{J.~L.} \bibnamefont{Rodriguez}},
  \bibinfo{author}{\bibfnamefont{S.~S.~M.} \bibnamefont{Wong}},
  \bibnamefont{and} \bibinfo{author}{\bibfnamefont{J.~K.} \bibnamefont{Tuli}},
  \bibinfo{journal}{Nuclear Data Sheets} \textbf{\bibinfo{volume}{84}},
  \bibinfo{pages}{487 } (\bibinfo{year}{1998}), ISSN \bibinfo{issn}{0090-3752},
  \urlprefix\url{http://www.sciencedirect.com/science/article/pii/S0090375298900151}.

\bibitem[{\citenamefont{Elekes et~al.}(2011)\citenamefont{Elekes, Timar, and
  Singh}}]{Elekes2011}
\bibinfo{author}{\bibfnamefont{Z.}~\bibnamefont{Elekes}},
  \bibinfo{author}{\bibfnamefont{J.}~\bibnamefont{Timar}}, \bibnamefont{and}
  \bibinfo{author}{\bibfnamefont{B.}~\bibnamefont{Singh}},
  \bibinfo{journal}{Nucl. Data Sheets} \textbf{\bibinfo{volume}{112}},
  \bibinfo{pages}{1} (\bibinfo{year}{2011}).

\bibitem[{log()}]{logft_nndc}
\emph{\bibinfo{title}{{Logft} log({\it ft}) calculator}},
  \bibinfo{howpublished}{\url{https://www.nndc.bnl.gov/logft/}},
  \bibinfo{note}{accessed: 2019-03-02}.

\bibitem[{\citenamefont{Wang et~al.}(2017)\citenamefont{Wang, Audi, Kondev,
  Huang, Naimi, and Xu}}]{ame2016}
\bibinfo{author}{\bibfnamefont{M.}~\bibnamefont{Wang}},
  \bibinfo{author}{\bibfnamefont{G.}~\bibnamefont{Audi}},
  \bibinfo{author}{\bibfnamefont{F.}~\bibnamefont{Kondev}},
  \bibinfo{author}{\bibfnamefont{W.}~\bibnamefont{Huang}},
  \bibinfo{author}{\bibfnamefont{S.}~\bibnamefont{Naimi}}, \bibnamefont{and}
  \bibinfo{author}{\bibfnamefont{X.}~\bibnamefont{Xu}}, \bibinfo{journal}{Chin.
  Phys. C} \textbf{\bibinfo{volume}{41}}, \bibinfo{pages}{030003}
  (\bibinfo{year}{2017}).

\bibitem[{\citenamefont{Hardy and Towner}(2020)}]{Hardy2020}
\bibinfo{author}{\bibfnamefont{J.~C.} \bibnamefont{Hardy}} \bibnamefont{and}
  \bibinfo{author}{\bibfnamefont{I.~S.} \bibnamefont{Towner}},
  \bibinfo{journal}{Phys. Rev. C} \textbf{\bibinfo{volume}{102}},
  \bibinfo{pages}{045501} (\bibinfo{year}{2020}),
  \urlprefix\url{https://link.aps.org/doi/10.1103/PhysRevC.102.045501}.

\bibitem[{\citenamefont{Barnes et~al.}(1965)\citenamefont{Barnes, Bockelman,
  Hansen, and Sperduto}}]{Barnes65}
\bibinfo{author}{\bibfnamefont{P.~D.} \bibnamefont{Barnes}},
  \bibinfo{author}{\bibfnamefont{C.~K.} \bibnamefont{Bockelman}},
  \bibinfo{author}{\bibfnamefont{O.}~\bibnamefont{Hansen}}, \bibnamefont{and}
  \bibinfo{author}{\bibfnamefont{A.}~\bibnamefont{Sperduto}},
  \bibinfo{journal}{Phys. Rev.} \textbf{\bibinfo{volume}{140}},
  \bibinfo{pages}{B42} (\bibinfo{year}{1965}),
  \urlprefix\url{https://link.aps.org/doi/10.1103/PhysRev.140.B42}.

\bibitem[{\citenamefont{Sona et~al.}(1984)\citenamefont{Sona, Mando, and
  Taccetti}}]{Sona84}
\bibinfo{author}{\bibfnamefont{P.}~\bibnamefont{Sona}},
  \bibinfo{author}{\bibfnamefont{P.~A.} \bibnamefont{Mando}}, \bibnamefont{and}
  \bibinfo{author}{\bibfnamefont{N.}~\bibnamefont{Taccetti}},
  \bibinfo{journal}{J. Phys. G: Nucl. Phys.} \textbf{\bibinfo{volume}{10}},
  \bibinfo{pages}{833 } (\bibinfo{year}{1984}),
  \urlprefix\url{https://iopscience.iop.org/article/10.1088/0305-4616/10/6/016}.

\bibitem[{\citenamefont{Brown and Rae}(2014)}]{Brown2014}
\bibinfo{author}{\bibfnamefont{B.~A.} \bibnamefont{Brown}} \bibnamefont{and}
  \bibinfo{author}{\bibfnamefont{W.~D.~M.} \bibnamefont{Rae}},
  \bibinfo{journal}{Nucl. Data Sheets} \textbf{\bibinfo{volume}{120}},
  \bibinfo{pages}{115} (\bibinfo{year}{2014}).

\bibitem[{\citenamefont{Honma et~al.}(2005)\citenamefont{Honma, Otsuka, Brown,
  and Mizusaki}}]{Honma2005}
\bibinfo{author}{\bibfnamefont{M.}~\bibnamefont{Honma}},
  \bibinfo{author}{\bibfnamefont{T.}~\bibnamefont{Otsuka}},
  \bibinfo{author}{\bibfnamefont{B.~A.} \bibnamefont{Brown}}, \bibnamefont{and}
  \bibinfo{author}{\bibfnamefont{T.}~\bibnamefont{Mizusaki}},
  \bibinfo{journal}{Eur. Phys. J. A} \textbf{\bibinfo{volume}{25 (Suppl. 1)}},
  \bibinfo{pages}{499} (\bibinfo{year}{2005}).

\bibitem[{\citenamefont{Tsukiyama et~al.}(2012)\citenamefont{Tsukiyama, Bogner,
  and Schwenk}}]{Tsuk12SM}
\bibinfo{author}{\bibfnamefont{K.}~\bibnamefont{Tsukiyama}},
  \bibinfo{author}{\bibfnamefont{S.~K.} \bibnamefont{Bogner}},
  \bibnamefont{and} \bibinfo{author}{\bibfnamefont{A.}~\bibnamefont{Schwenk}},
  \bibinfo{journal}{Phys. Rev. C} \textbf{\bibinfo{volume}{85}},
  \bibinfo{pages}{061304(R)} (\bibinfo{year}{2012}).

\bibitem[{\citenamefont{Stroberg et~al.}(2019)\citenamefont{Stroberg, Hergert,
  Bogner, and Holt}}]{Stroberg:2019mxo}
\bibinfo{author}{\bibfnamefont{S.~R.} \bibnamefont{Stroberg}},
  \bibinfo{author}{\bibfnamefont{H.}~\bibnamefont{Hergert}},
  \bibinfo{author}{\bibfnamefont{S.~K.} \bibnamefont{Bogner}},
  \bibnamefont{and} \bibinfo{author}{\bibfnamefont{J.~D.} \bibnamefont{Holt}},
  \bibinfo{journal}{Annual Review of Nuclear and Particle Science}
  \textbf{\bibinfo{volume}{69}}, \bibinfo{pages}{307} (\bibinfo{year}{2019}),
  \eprint{https://doi.org/10.1146/annurev-nucl-101917-021120},
  \urlprefix\url{https://doi.org/10.1146/annurev-nucl-101917-021120}.

\bibitem[{\citenamefont{Bogner et~al.}(2014)\citenamefont{Bogner, Hergert,
  Holt, Schwenk, Binder, Calci, Langhammer, and Roth}}]{Bogn14SM}
\bibinfo{author}{\bibfnamefont{S.~K.} \bibnamefont{Bogner}},
  \bibinfo{author}{\bibfnamefont{H.}~\bibnamefont{Hergert}},
  \bibinfo{author}{\bibfnamefont{J.~D.} \bibnamefont{Holt}},
  \bibinfo{author}{\bibfnamefont{A.}~\bibnamefont{Schwenk}},
  \bibinfo{author}{\bibfnamefont{S.}~\bibnamefont{Binder}},
  \bibinfo{author}{\bibfnamefont{A.}~\bibnamefont{Calci}},
  \bibinfo{author}{\bibfnamefont{J.}~\bibnamefont{Langhammer}},
  \bibnamefont{and} \bibinfo{author}{\bibfnamefont{R.}~\bibnamefont{Roth}},
  \bibinfo{journal}{Phys. Rev. Lett.} \textbf{\bibinfo{volume}{113}},
  \bibinfo{pages}{142501} (\bibinfo{year}{2014}).

\bibitem[{\citenamefont{Stroberg et~al.}(2017)\citenamefont{Stroberg, Calci,
  Hergert, Holt, Bogner, Roth, and Schwenk}}]{Stro17ENO}
\bibinfo{author}{\bibfnamefont{S.~R.} \bibnamefont{Stroberg}},
  \bibinfo{author}{\bibfnamefont{A.}~\bibnamefont{Calci}},
  \bibinfo{author}{\bibfnamefont{H.}~\bibnamefont{Hergert}},
  \bibinfo{author}{\bibfnamefont{J.~D.} \bibnamefont{Holt}},
  \bibinfo{author}{\bibfnamefont{S.~K.} \bibnamefont{Bogner}},
  \bibinfo{author}{\bibfnamefont{R.}~\bibnamefont{Roth}}, \bibnamefont{and}
  \bibinfo{author}{\bibfnamefont{A.}~\bibnamefont{Schwenk}},
  \bibinfo{journal}{Phys. Rev. Lett.} \textbf{\bibinfo{volume}{118}},
  \bibinfo{pages}{032502} (\bibinfo{year}{2017}).

\bibitem[{\citenamefont{Stroberg et~al.}(2016)\citenamefont{Stroberg, Hergert,
  Holt, Bogner, and Schwenk}}]{Stro16TNO}
\bibinfo{author}{\bibfnamefont{S.~R.} \bibnamefont{Stroberg}},
  \bibinfo{author}{\bibfnamefont{H.}~\bibnamefont{Hergert}},
  \bibinfo{author}{\bibfnamefont{J.~D.} \bibnamefont{Holt}},
  \bibinfo{author}{\bibfnamefont{S.~K.} \bibnamefont{Bogner}},
  \bibnamefont{and} \bibinfo{author}{\bibfnamefont{A.}~\bibnamefont{Schwenk}},
  \bibinfo{journal}{Phys. Rev. C} \textbf{\bibinfo{volume}{93}},
  \bibinfo{pages}{051301(R)} (\bibinfo{year}{2016}).

\bibitem[{\citenamefont{Hebeler et~al.}(2011)\citenamefont{Hebeler, Bogner,
  Furnstahl, Nogga, and Schwenk}}]{Hebe11fits}
\bibinfo{author}{\bibfnamefont{K.}~\bibnamefont{Hebeler}},
  \bibinfo{author}{\bibfnamefont{S.~K.} \bibnamefont{Bogner}},
  \bibinfo{author}{\bibfnamefont{R.~J.} \bibnamefont{Furnstahl}},
  \bibinfo{author}{\bibfnamefont{A.}~\bibnamefont{Nogga}}, \bibnamefont{and}
  \bibinfo{author}{\bibfnamefont{A.}~\bibnamefont{Schwenk}},
  \bibinfo{journal}{Phys. Rev. C} \textbf{\bibinfo{volume}{83}},
  \bibinfo{pages}{031301(R)} (\bibinfo{year}{2011}).

\bibitem[{\citenamefont{Simonis et~al.}(2016)\citenamefont{Simonis, Hebeler,
  Holt, Men{\'e}ndez, and Schwenk}}]{Simo16unc}
\bibinfo{author}{\bibfnamefont{J.}~\bibnamefont{Simonis}},
  \bibinfo{author}{\bibfnamefont{K.}~\bibnamefont{Hebeler}},
  \bibinfo{author}{\bibfnamefont{J.~D.} \bibnamefont{Holt}},
  \bibinfo{author}{\bibfnamefont{J.}~\bibnamefont{Men{\'e}ndez}},
  \bibnamefont{and} \bibinfo{author}{\bibfnamefont{A.}~\bibnamefont{Schwenk}},
  \bibinfo{journal}{Phys. Rev. C} \textbf{\bibinfo{volume}{93}},
  \bibinfo{pages}{011302(R)} (\bibinfo{year}{2016}).

\bibitem[{\citenamefont{Simonis et~al.}(2017)\citenamefont{Simonis, Stroberg,
  Hebeler, Holt, and Schwenk}}]{Simo17SatFinNuc}
\bibinfo{author}{\bibfnamefont{J.}~\bibnamefont{Simonis}},
  \bibinfo{author}{\bibfnamefont{S.~R.} \bibnamefont{Stroberg}},
  \bibinfo{author}{\bibfnamefont{K.}~\bibnamefont{Hebeler}},
  \bibinfo{author}{\bibfnamefont{J.~D.} \bibnamefont{Holt}}, \bibnamefont{and}
  \bibinfo{author}{\bibfnamefont{A.}~\bibnamefont{Schwenk}},
  \bibinfo{journal}{Phys. Rev. C} \textbf{\bibinfo{volume}{96}},
  \bibinfo{pages}{014303} (\bibinfo{year}{2017}).

\bibitem[{\citenamefont{Morris et~al.}(2015)\citenamefont{Morris, Parzuchowski,
  and Bogner}}]{Morr15Magnus}
\bibinfo{author}{\bibfnamefont{T.~D.} \bibnamefont{Morris}},
  \bibinfo{author}{\bibfnamefont{N.~M.} \bibnamefont{Parzuchowski}},
  \bibnamefont{and} \bibinfo{author}{\bibfnamefont{S.~K.}
  \bibnamefont{Bogner}}, \bibinfo{journal}{Phys. Rev. C}
  \textbf{\bibinfo{volume}{92}}, \bibinfo{pages}{034331}
  (\bibinfo{year}{2015}),
  \urlprefix\url{https://link.aps.org/doi/10.1103/PhysRevC.92.034331}.

\bibitem[{\citenamefont{Hergert et~al.}(2016)\citenamefont{Hergert, Bogner,
  Morris, Schwenk, and Tsukiyama}}]{Herg16PR}
\bibinfo{author}{\bibfnamefont{H.}~\bibnamefont{Hergert}},
  \bibinfo{author}{\bibfnamefont{S.~K.} \bibnamefont{Bogner}},
  \bibinfo{author}{\bibfnamefont{T.~D.} \bibnamefont{Morris}},
  \bibinfo{author}{\bibfnamefont{A.}~\bibnamefont{Schwenk}}, \bibnamefont{and}
  \bibinfo{author}{\bibfnamefont{K.}~\bibnamefont{Tsukiyama}},
  \bibinfo{journal}{Phys. Rep.} \textbf{\bibinfo{volume}{621}},
  \bibinfo{pages}{165} (\bibinfo{year}{2016}).

\bibitem[{\citenamefont{Parzuchowski et~al.}(2017)\citenamefont{Parzuchowski,
  Stroberg, Navratil, Hergert, and Bogner}}]{Parz:prc2017}
\bibinfo{author}{\bibfnamefont{N.~M.} \bibnamefont{Parzuchowski}},
  \bibinfo{author}{\bibfnamefont{S.~R.} \bibnamefont{Stroberg}},
  \bibinfo{author}{\bibfnamefont{P.}~\bibnamefont{Navratil}},
  \bibinfo{author}{\bibfnamefont{H.}~\bibnamefont{Hergert}}, \bibnamefont{and}
  \bibinfo{author}{\bibfnamefont{S.~K.} \bibnamefont{Bogner}},
  \bibinfo{journal}{Phys. Rev. C} \textbf{\bibinfo{volume}{96}},
  \bibinfo{pages}{034324} (\bibinfo{year}{2017}).

\bibitem[{\citenamefont{Henderson et~al.}(2018)}]{Henderson:2017dqc}
\bibinfo{author}{\bibfnamefont{J.}~\bibnamefont{Henderson}}
  \bibnamefont{et~al.}, \bibinfo{journal}{Phys. Lett.}
  \textbf{\bibinfo{volume}{B782}}, \bibinfo{pages}{468} (\bibinfo{year}{2018}).

\bibitem[{\citenamefont{Caprio}(2005)}]{Caprio2005}
\bibinfo{author}{\bibfnamefont{M.~A.} \bibnamefont{Caprio}},
  \bibinfo{journal}{Comput. Phys. Commun.} \textbf{\bibinfo{volume}{171}},
  \bibinfo{pages}{107 } (\bibinfo{year}{2005}),
  \urlprefix\url{http://scidraw.nd.edu}.

\end{thebibliography}

\end{document}